\pdfoutput=1

\documentclass[fleqn,usenatbib]{mnras}

\usepackage[T1]{fontenc}

\DeclareRobustCommand{\VAN}[3]{#2}
\let\VANthebibliography\thebibliography
\def\thebibliography{\DeclareRobustCommand{\VAN}[3]{##3}\VANthebibliography}

\usepackage{graphicx}	
\usepackage{amsmath}	
\usepackage{amssymb}	

\usepackage{newtxtext,newtxmath}

\title[Lens modelling: BNNs and parametric fitting]{Strong lens modelling: comparing and combining Bayesian neural networks and parametric profile fitting}

\author[J. Pearson et al.]{
James Pearson,$^{1}$\thanks{E-mail: james.pearson@nottingham.ac.uk}
Jacob Maresca,$^{1}$
Nan Li$^{1,2}$
and Simon Dye$^{1}$
\\
$^{1}$School of Physics and Astronomy, University of Nottingham, University Park, Nottingham, NG7 2RD, UK\\
$^{2}$National Astronomical Observatories of China, 20A Datun Road, Chaoyang District, Beijing, 100012, China\\
}

\date{Accepted 2021 May 22. Received 2021 May 3; in original form 2021 March 4}

\pubyear{2021}

\begin{document}
\label{firstpage}
\pagerange{\pageref{firstpage}--\pageref{lastpage}}
\maketitle

\begin{abstract}
The vast quantity of strong galaxy-galaxy gravitational lenses expected by future large-scale surveys necessitates the development of automated methods to efficiently model their mass profiles. For this purpose, we train an approximate Bayesian convolutional neural network (CNN) to predict mass profile parameters and associated uncertainties, and compare its accuracy to that of conventional parametric modelling for a range of increasingly complex lensing systems. 
These include standard smooth parametric density profiles, hydrodynamical EAGLE galaxies and the inclusion of foreground mass structures, combined with parametric sources and sources extracted from the Hubble Ultra Deep Field.
In addition, we also present a method for combining the CNN with traditional parametric density profile fitting in an automated fashion, where the CNN provides initial priors on the latter's parameters.
On average, the CNN achieved errors 19 $\pm$ 22 per cent lower than the traditional method's blind modelling.
The combination method instead achieved 27 $\pm$ 11 per cent lower errors over the blind modelling, reduced further to 37 $\pm$ 11 per cent when the priors also incorporated the CNN-predicted uncertainties, with errors also 17 $\pm$ 21 per cent lower than the CNN by itself.
While the CNN is undoubtedly the fastest modelling method, the combination of the two increases the speed of conventional fitting alone by factors of 1.73 and 1.19 with and without CNN-predicted uncertainties, respectively.
This, combined with greatly improved accuracy, highlights the benefits one can obtain through combining neural networks with conventional techniques in order to achieve an efficient automated modelling approach.
\end{abstract}

\begin{keywords}
gravitational lensing: strong -- galaxies: structure
\end{keywords}



\section{Introduction}
\label{sec:introduction}


The phenomenon of strong galaxy-galaxy lensing, whereby a foreground galaxy strongly lenses a background galaxy, provides a means of studying various physical properties of the Universe. Measurements of the observed distortion allow for modelling of the projected mass density profile of the foreground galaxy, which contains information on the dark matter content and substructure within the lens \citep{sonnenfeld2015sl2s,shu2017sloan,kung2018models}. Advancements have recently been made towards detecting this substructure \citep{vegetti2009bayesian,vegetti2014inference,hezaveh2016detection,bayer2018observational,brehmer2019mining,ritondale2019low}, with such properties aiding in galaxy evolution models \citep[e.g.][]{bolton2012boss}.

Lensing maintains the surface brightness of sources, but the resulting sheared and magnified images allow for the probing of high-redshift source populations, especially if their original surface brightness distributions can be reconstructed. With the addition of redshift measurements, this too can provide valuable information on galaxy evolution, and as such has received a recent surge in interest \citep[e.g.][]{dye2018modelling,lemon2018gravitationally,mcgreer2018bright,rubin2018discovery,salmon2018relics,sharda2018testing,shu2018sdss,sharon2019resolved,collett2020triple,khullar2020cool,inoue2020alma}. Reconstructing the unlensed morphology of a source is possible if the mass profile of the lens is well constrained \citep{warren2003semilinear,suyu2006bayesian,nightingale2018autolens,powell2021novel}, and allows for a more in-depth study of their properties, for example, their rotation curves \citep{dye2015revealing,geach2018magnified}.

The projected mass profiles of the foreground lenses, when combined with other methods such as galaxy rotation curves, may be used to obtain approximate 3D (deprojected) mass density profiles. Such profiles can be of use in testing General Relativity \citep{collett2018precise} and cosmological models \citep{eales2015h,krywult2017vimos,rana2017probing,davies2018galaxy,yang2020first,colacco2020probing,luo2020emergent,giani2020testing,birrer2020tdcosmo,shajib2020strides,wang2020cosmological}. Gravitational time delays and geometric path differences between images, paired with variable sources such as quasars, provide measurable time delays that can constrain the value of the Hubble constant ($H_0$) irrespective of the distances to or between the galaxies \citep[e.g.][]{suyu2017h0licow,bonvin2017h0licow,birrer2019h0licow,chen2018constraining,liao2019model,taubenberger2019hubble,birrer2020tdcosmo2,li2020impact,wong2020h0licow,denzel2021hubble}. 
Recently, there has been much work on extending this to gravitationally lensed supernovae, whose standardisable absolute brightnesses and well-understood light curves may provide far tighter constraints on $H_0$ \citep{collett2019model,oguri2019strong,foxley2020observing,bag2021unresolved,bayer2021holismokes}.
There remains significant tension in the predicted value of the Hubble constant between techniques focusing on early-universe and late-universe physics \citep{freedman2017cosmology,mortsell2018does,bengaly2020hubble,pandey2020model,vagnozzi2020new}, with the modelling of larger strong lens catalogues key to reducing the uncertainties in lensing estimates of $H_0$.


Surveys involved in identifying strong lenses have to date confirmed only hundreds, with most lying at low redshift. These include the Sloan Lens ACS (SLACS) survey \citep{bolton2006sloan}, the CFHTLS Strong Lensing Legacy Survey \citep[SL2S;][]{cabanac2007cfhtls}, the Sloan WFC Edge-on Late-type Lens Survey \citep[SWELLS;][]{treu2011swells}, the BOSS Emission-Line Lens Survey \citep[BELLS;][]{brownstein2011boss} and lenses found in the Dark Energy Survey \citep{dark2005dark}. Upcoming surveys are expected to remedy this, producing billions of galaxy images containing tens of thousands of strong lensing systems \citep{collett2015population}. These include the European Space Agency's \textit{Euclid} telescope \citep{laureijs2011euclid}, and the Legacy Survey of Space and Time (LSST) which will use the ground-based Vera C. Rubin Observatory (formerly the Large Synoptic Survey Telescope \citep[LSST;][]{ivezic2008large}).
\textit{Euclid} is due to launch in 2022 with the primary aim of measuring the acceleration of the Universe up to a redshift of $z=2$ to study dark matter and dark energy, covering 15,000 deg$^{2}$ over its six year mission. LSST will begin science operations in 2023, covering around 18,000 deg$^{2}$ in six bands ($u$, $g$, $r$, $i$, $z$, $y$) repeatedly over ten years, also studying dark matter and dark energy.


As a result, multiple automated methods have been developed for rapidly and accurately identifying strong gravitational lenses.
Convolutional Neural Networks (CNNs) have seen extensive use in this area \citep{jacobs2017finding,lanusse2017cmu,petrillo2017finding,schaefer2018deep,davies2019using,jacobs2019extended,metcalf2019strong,canameras2020holismokes,he2020deep,li2020new,huang2021discovering}, along with other machine learning methods \citep{cheng2020identifying}, as these require neither the spectroscopic data nor arbitrary geometric measurements often employed by other techniques \citep[e.g.][]{avestruz2019automated,bom2017neural,ostrovski2017discovery,talbot2020completed}. CNNs are a subset of deep neural networks that have in recent years become popular for handling large amounts of data, such as rapid feature extraction and classification of images, and have seen a wide range of applications in astronomy \citep[e.g.][]{paillassa2020maximask,schuldt2020photometric,tohill2020measuring,wu2020predicting}. Such networks have been shown to be very effective at correctly identifying thousands of lenses purely from images, and are able to do so extremely quickly. However, this first requires the CNN to be trained on many tens of thousands to hundreds of thousands of images; with so few real images of gravitational lenses, these training sets must be simulated instead.
While lens detection has also seen the application of citizen science \citep{more2015space,sonnenfeld2020survey}, there may be little overlap between lenses identified by citizen science and those by machine learning \citep{knabel2020galaxy}, highlighting the need for multiple approaches in order to obtain the most complete sample.


Following identification of a lens system, mass modelling is typically performed using parametric techniques. These obtain the lens mass model parameters that best fit the observed image. Different techniques have been developed for this purpose, for example those that involve pixellated grids to reconstruct sources \citep{warren2003semilinear,vegetti2009bayesian,nightingale2018autolens} or the use of shapelets \citep{birrer2015gravitational,birrer2018lenstronomy}. Modelling can often require time-consuming processes including point spread function estimation, removal of the lens galaxy light and image masking prior to modelling; to circumvent some of this initial processing these techniques now simultaneously fit both the lens galaxy light and mass profile but at the expense of an even slower modelling speed.

With CNNs successfully used for detecting lenses, they have since been shown to provide a promising alternative method of lens modelling. 
\cite{hezaveh2017fast} demonstrated this for the first time, training a combination of networks to predict lens mass model parameters and applying them to simulated and real Hubble Space Telescope (HST) images. A method of obtaining uncertainties on these predictions was presented by \cite{levasseur2017uncertainties}, with the CNN now an approximate Bayesian neural network (BNN). While the initial training took multiple days on a GPU machine, when applied to test images they reported an increase in lens modelling speed of several orders of magnitude compared to parametric methods, demonstrating the potential application of CNNs for this purpose. This was later extended for application to interferometric observations \citep{morningstar2018analyzing}, along with the demonstration of machine learning to additionally reconstruct the background source from CNN-predicted parameters \cite{morningstar2019data}.

Since then, CNNs and similar networks have seen much use in lens modelling, including source reconstruction \citep{chianese2019differentiable,chianese2020differentiable}, redshift and lens velocity dispersion estimation \citep{bom2019deep} and the detection and modelling of dark matter substructure \citep{brehmer2019mining,alexander2020deep,lin2020hunting,rivero2020direct,varma2020dark,vernardos2020quantifying}. Recent work by \cite{maresca2020auto} showed how CNNs could easily identify unphysical source reconstructions outputted by semi-analytic parametric modelling methods, allowing for an automated approach to dealing with incorrect models. With regard to using CNNs to obtain parametric lens mass profile parameters, \cite{schuldt2021holismokes} focused on ground-based imaging, leaving in foreground lens light and making use of four filters to distinguish sources composed of Hubble Ultra-Deep Field \citep[HUDF;][]{beckwith2006hubble} galaxies before examining how well the CNN-predicted models translated into predicting time delays and image positions. \cite{park2021large} applied an approximate BNN to the modelling of time-delay lenses consisting of lensed AGN, combining results with simulated time delays for $H_0$ inference across hundreds of such lenses. 
\cite{wagner2021hierarchical} presented a hierarchical inference framework for such BNN lens modelling to avoid biases introduced by differences between training and test data sets.
And \cite{madireddy2019modular} presented a pipeline for both lens detection and modelling, including denoising and deblending (removing lens light), using multiple different deep neural networks in a modular fashion that provided informative latent spaces and uncertainties at each stage.


In \cite{pearson2019use}, we investigated the practicalities of using CNNs for lens modelling. Firstly, we examined how such networks cope when trained to model the lens mass profile without prior removal of lens light, including an assessment of the impact of assumed mass and light alignment during training. Secondly, we quantified the gain in accuracy through using multiband imaging, and compared both of these results to the case of modelling the mass profile when the lens light was removed. The data sets used in \cite{pearson2019use} were simulated to match the imaging characteristics of both the \textit{Euclid} telescope and LSST's Vera C. Rubin Observatory, in preparation for the large quantity of lenses that will be observed by these large-scale surveys.


In this paper, we seek to answer the the following: How well can we expect CNNs (now approximate BNNs) to obtain lens mass model parameters for increasingly realistic images? How do they compare to conventional parameter-fitting techniques? 
What benefits arise from a combination of machine learning and parametric modelling methods and can the disadvantages of one be counteracted by the other?
While trained neural networks are much faster at modelling, they are limited by the quality of their training data; mismatches between these and real observations can make neural networks less reliable than their conventional parameter-fitting counterparts.
For this work, we focus on the imaging characteristics of the \textit{Euclid} telescope only to avoid an excess of results, and compare the CNN to the semilinear inversion technique of \textsc{PyAutoLens} \citep{nightingale2018autolens} as well as a combination of the two.
These modelling methods are tested on a range of increasingly complex lensing systems, starting with smooth singular isothermal ellipsoid (SIE) lenses and parametric sources before incorporating real Hubble Ultra-Deep Field \citep[HUDF;][]{beckwith2006hubble} source galaxies and CosmoDC2 \citep{korytov2019cosmodc2} line-of-sight (LOS) structure, and replacing the foreground lenses with more complex power law mass profiles or galaxies from the Evolution and Assembly of Galaxies and their Environments (EAGLE) cosmological hydrodynamical simulation suite \citep{crain2015eagle,schaye2015eagle}.

The paper is organised as follows: Section \ref{sec:grav_lens_sims} gives the details of the simulations used to produce the image data sets used for training and testing. Section \ref{sec:lens_modelling_methodology} provides an overview of the parameter-fitting methods used in this paper: \textsc{PyAutoLens} and the CNN, including the network's architecture, uncertainties and training. Section \ref{sec:results} presents and compares the results from testing these methods on multiple image data sets of increasing complexity, from simple parametric lenses and sources to hydrodynamical EAGLE galaxy lenses with real HUDF sources and LOS structure. The results are discussed in Section \ref{sec:discussion} along with a conclusion of this work in Section \ref{sec:summary}.

\section{Gravitational Lens Simulation}
\label{sec:grav_lens_sims}

In this section, we describe gravitational lens image simulations used throughout this work. As the CNN used here is a form of supervised machine learning, a large data set of images is required to train the network before it can be applied to test images. 
This training set contained parametric lens mass models and parametric source light profiles, while the multiple test sets contained various profiles, beginning parametrically before replacing these with more realistic lenses and sources.
Section \ref{subsec:training_sims} and Section \ref{subsec:testing_sims} give details of the simulated data sets used for training the network and testing the modelling methods, respectively.

\subsection{Training set}
\label{subsec:training_sims}

The data set of 100,000 images used to train the CNN was generated following the simulation method detailed in \cite{pearson2019use}.
As the majority of strong gravitational lenses are early-type galaxies, for the lens mass profile we adopt the SIE model \citep{kormann1994isothermal,keeton2001catalog} commonly used as a good fit for strong lens profiles, as well as its more general form, the power law ellipsoid \citep{tessore2015elliptical}. The former has three main mass model parameters (excluding the coordinates of the lens): Einstein radius, orientation and axis ratio. The latter has an additional parameter, the power law slope, that allows for increased flexibility but increases the complexity of the parameter space when modelling. 
As such, the training set was generated twice, once with each mass model, in order to produce one CNN trained to predict three SIE parameters and another for the four power law parameters. The CNNs were otherwise identical in their training and architecture (see Section \ref{subsec:conv_neural_networks}). 
To ensure sufficient training across parameter space, the data sets contained lenses with uniformly distributed parameters selected from the following ranges: 0.2-3.2 arcseconds, 0-180$^{\circ}$, 0.4-1.0, and 1.5-2.5 for Einstein radii, orientations, axis ratios and slopes, respectively.
The range of Einstein radii was motivated by the distributions of lenses expected to be obtained by the \textit{Euclid} survey \citep{collett2015population}.
The lenses were assumed to be correctly centred, with central positions following a normal distribution about the image centre with a standard deviation of one pixel.

\begin{figure*}
    \includegraphics[width=17cm]{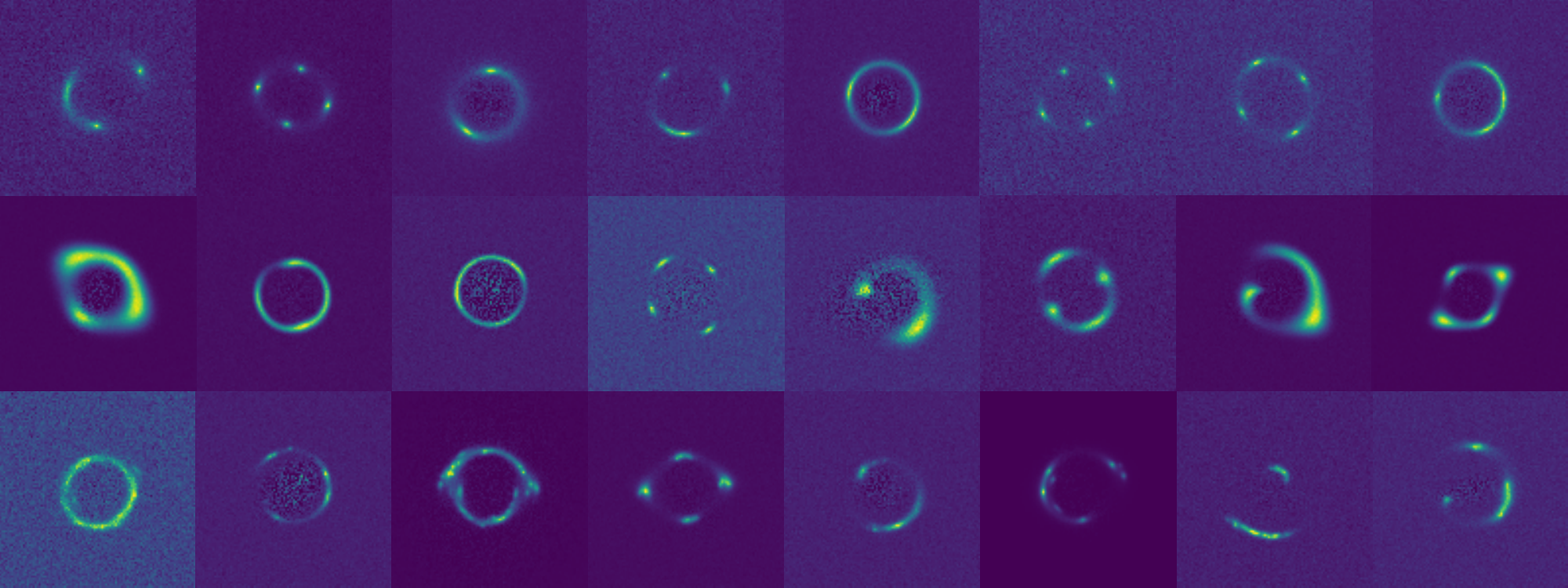}
    \caption{Examples of the simulated images used to train the neural network. Top and middle rows contain S\'ersic and Gaussian profile sources, respectively. The bottom row contains complex sources made up of multiple S\'ersic profiles. The data set was simulated to have the expected imaging characteristics of the \textit{Euclid} telescope's VIS filter.}
    \label{fig:training_imgs}
\end{figure*}

For the comparisons made in this work, all images were simulated with the characteristics expected of the \textit{Euclid} telescope's VIS filter, adopting its native pixel scale \citep[0.1 arcsec pixel$^{-1}$;][]{racca2016euclid} and convolving the images with a Gaussian point spread function with a full width at half-maximum of 0.17 arcseconds. The postage stamp images were fixed at $100 \times 100$ pixels based on the distribution of expected Einstein radii.
In addition to the sources with parametric S\'ersic profiles used in \cite{pearson2019use}, the training data set contained sources with Gaussian profiles as well as complex sources made up of multiple S\'ersic profiles, in order to cover a wider range of morphologies.
These S\'ersic or Gaussian profiles were generated with variable S\'ersic indices and variances respectively, each making up 25 per cent of the training set while the remaining 50 per cent contained multiple S\'ersic profiles.
Example training images are shown in Fig. \ref{fig:training_imgs}.
To ensure the network would not only be trained to cope with images containing highly magnified sources, all source centroid positions were determined randomly from a uniform distribution within the Einstein radius of the lens.

Based on trial data sets and work by \cite{collett2015population}, lens and source redshifts were drawn from uniform distributions with upper limits of $z_{\rm lens}=2$ and $z_{\rm source}=6$, and multiple detection criteria were implemented to ensure strong gravitational lenses were produced. Requirements included that the centre of the source must be multiply-imaged, with the counter image resolved, and that there was sufficient magnification ($\mu_{\rm TOT} > 3$), tangential shearing ($\mu_{\rm TOT} R_{\rm e, source} >$ seeing) and signal-to-noise ratio (SNR > 20).

The sky background for \textit{Euclid} VIS \citep{collett2015population,euclid2015python} was added along with shot noise, the expected read noise (five electrons per readout) and dark current \citep[two electrons per pixel per second;][]{radeka2009lsst}.
In addition to the lensed source, such images initially contained light from the foreground lens, before subtracting the true light profile convolved with the point spread function to leave only shot noise residuals.

\subsection{Test sets}
\label{subsec:testing_sims}

The 1000-image data sets used for testing the CNN and \textsc{PyAutoLens} were generated with the Pipeline for Images of Cosmological Strong Lensing \citep[PICS;][]{li2016pics}. We made use of their selection of Hubble Ultra-Deep Field \citep[HUDF;][]{beckwith2006hubble} galaxies for high-resolution, high-redshift sources, although we did not attempt to deconvolve these sources for this work. 
These single-band test images had the same size and imaging characteristics as the training set, but did not contain any shot noise residuals of the lenses as the errors introduced by lens removal depend on the technique used and are not the focus of this work.
The magnitudes and angular sizes of the HUDF source galaxies were incorporated into the images, including the data set containing parametric (S\'ersic and Gaussian) light profiles.
As such, lens and source redshifts had the sole effect of changing the Einstein radii, so for the purposes of the simulation redshifts were taken to be $z_{lens}=0.5$ and $z_{source}=2.0$, respectively, with Einstein radii set manually.

\begin{figure*}
    \includegraphics[width=17cm]{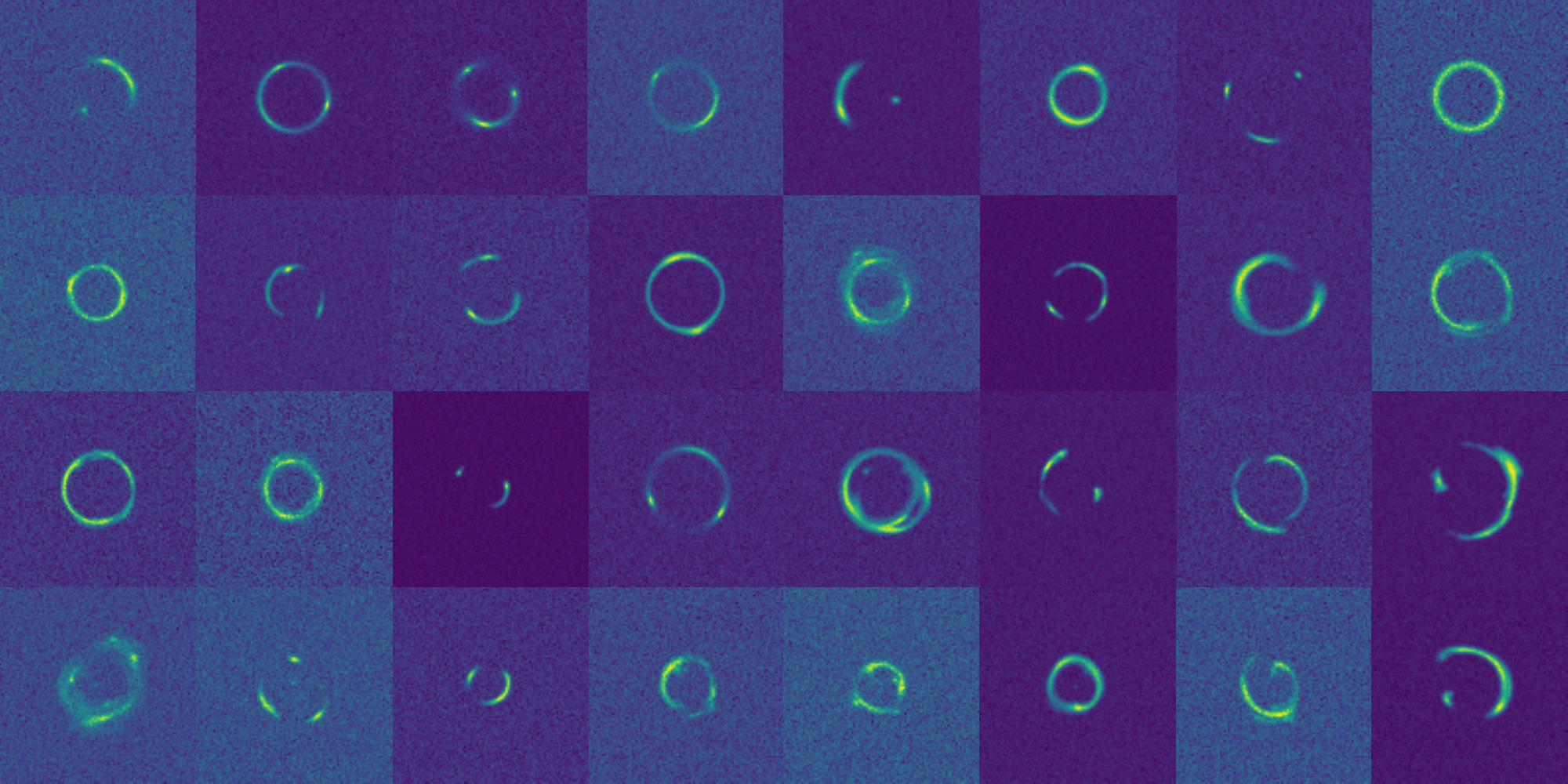}
    \caption{Examples of the simulated images produced from PICS software, used to test the different mass modelling methods. Top row: examples containing SIE lenses with parametric source profiles, with four S\'ersic sources followed by four Gaussian sources. Second row: examples containing SIE lenses with HUDF sources, with the last four containing LOS structure. Third row: examples containing power law lenses with HUDF sources, with the last four containing LOS structure. Bottom row: examples containing EAGLE galaxy lenses with HUDF sources, with the last four containing LOS structure. The data sets were simulated to have the expected imaging characteristics of the \textit{Euclid} telescope's VIS filter.}
    \label{fig:test_imgs}
\end{figure*}

PICS also allowed for the inclusion of LOS structure through full ray-tracing of small light cones covering lens caustics, with light cones from the semi-analytic model CosmoDC2 \citep{korytov2019cosmodc2} based on the Outer Rim cosmological \textit{N}-body simulation \citep{heitmann2019outer}. One thousand of these small light cones were selected from within the field of view of the semi-analytic catalogue, each containing a primary lens often residing in a galaxy group, with up to a hundred LOS galaxies between the observer and source. 
Test data sets were produced for a range of increasingly complex lensing systems, examples of which are shown in Fig. \ref{fig:test_imgs}:
\begin{itemize}
    \item SIE lenses + simple parametric source profiles
    \item SIE lenses + HUDF sources, with and without LOS structure
    \item Power law lenses + HUDF sources, with and without LOS structure
    \item EAGLE galaxy lenses + HUDF sources, with and without LOS structure
\end{itemize}

For the SIE lens mass profiles, their parameter distributions followed those determined by PICS to be the main halos in each CosmoDC2 light cone, with resulting Einstein radii, orientations and axis ratios distributed within the ranges 1.3-2.6 arcseconds, 0-180$^{\circ}$ and 0.4-1.0, respectively. The power law lenses used the same distribution of lens parameters, with the addition of slope values randomly selected from a uniform distribution in the range 1.5-2.5, in line with the CNN training data. These main halo lenses were on average more massive, and therefore produced larger Einstein radii, than the majority of those expected by the \textit{Euclid} survey. However, for this work these masses were not changed from the CosmoDC2 values as, where possible, we wished to maintain the halo masses and positions along the LOS for realistic ray-tracing through the light cones.

The EAGLE galaxy lenses were selected from the EAGLE galaxy catalogue by calculating the Einstein radius of each galaxy from their halo mass (assuming redshifts of $z_{lens}=0.5$ and $z_{source}=2.0$) and selecting one thousand of those with Einstein radii within the range 0.5-5.0 arcseconds. This resulted in a sample of galaxies whose Einstein radii were naturally distributed between 0.5 and 3.2 arcseconds.
The parameters used as the true values for the results in Section \ref{subsec:eagle_hudf} therefore followed different distributions than those of the CosmoDC2 halos, although orientations and axis ratios were limited to the same ranges. They were obtained using PICS by performing a least squares fit to their known convergence maps (computed from the EAGLE particle data), weighted by the square root of these maps, and supplying SIE mass profiles as the curve-fitting function. The same process was also used to obtain parameters for the data sets containing LOS structure, fitting either an SIE or power law mass profile.

While the CNN and \textsc{PyAutoLens} only examine the observed images, the inclusion of LOS structure increases the observed Einstein radius compared to that of a solitary foreground lens. To eliminate such biases in the results, rather than using the Einstein radii of the foreground lenses as the true values, we instead measured the observed Einstein radii of each lensing system using the same convergence map fitting process as discussed previously.

\section{Lens modelling methodology}
\label{sec:lens_modelling_methodology}

This section details the techniques used to model the simulated lenses. These involve the conventional parametric parameter-fitting approach of \textsc{PyAutoLens}, the CNN, and the combination of the two. Details of \textsc{PyAutoLens} are given in Section \ref{subsec:pyautolens} while the architecture, uncertainty estimation and training of the CNN are given in Section \ref{subsec:conv_neural_networks}, followed by the combination method in Section \ref{subsec:pyal_plus_cnn}.

\subsection{\textsc{PyAutoLens}}
\label{subsec:pyautolens}

To compare the CNN with conventional parameter-fitting, we made use of the semilinear inversion method of \textsc{PyAutoLens}\footnote{\url{https://github.com/Jammy2211/PyAutoLens}} \citep{nightingale2018autolens}.
This software simultaneously models the foreground lens and background source for strongly-lensed systems in a Bayesian framework, and can do so using a range of configuration options including modelling on a square or adaptive Voronoi grid, regularisation, and applying different source plane weightings such as weighting by source brightness. 
The result of modelling is a $\chi^2$ statistic of the differences between the observed and model images, used to calculate the Bayesian evidence. By repeatedly updating the parameters, this evidence is then maximised over a series of modelling iterations in an attempt to obtain the set of parameter values that correspond to the global maximum (and hence the global minimum in $\chi^2$ space).
For this work, the background source was reconstructed on a Voronoi grid that adapted to the lens magnification with constant regularisation. The inversion process was initialised using the input priors on the mass profile, a regularisation prior, and a prior controlling the number of pixels in the source reconstruction.
Each pixel in the image plane was sub-gridded into 4 sub-pixels for ray tracing calculations, with images masked by fitting an annulus to significant image pixels; for more information see Section 3.2 of \cite{maresca2020auto}.
For an optimiser, \textsc{PyAutoLens} used the \textsc{MultiNest} Bayesian inference tool \citep{feroz2009multinest} for Markov chain Monte Carlo (MCMC) sampling of parameter space, run in constant efficiency mode with a sampling efficiency of 0.4.

\textsc{PyAutoLens} requires priors for the mass model parameters which are normally set manually by human inspection of the lensed image, with the widths of the bounds of the priors set relatively arbitrarily.
Automated modelling was required for this work, and so \textsc{PyAutoLens} was initialised using priors set either blindly or using CNN-predicted values (see Section \ref{subsec:pyal_plus_cnn}). For the former, before modelling, priors for Einstein radii were approximated by fitting a circle to image pixels above a 3-sigma threshold, using uniform weights. This eased the fitting process while still leaving it fully automated, so remained a fair comparison to the other automated modelling methods.

\subsection{Convolutional neural networks}
\label{subsec:conv_neural_networks}

The following section details the CNN used throughout this work. Its architecture, along with an overview of how such networks work, is given in Section \ref{subsec:cnn_architecture}. Section \ref{subsec:cnn_uncertainties} details how uncertainties were obtained on the CNN's predicted values in order for it to become an approximate BNN, and Section \ref{subsec:cnn_training} discusses how the network was trained.

\subsubsection{Architecture}
\label{subsec:cnn_architecture}

\begin{figure*}
    \includegraphics[width=17cm]{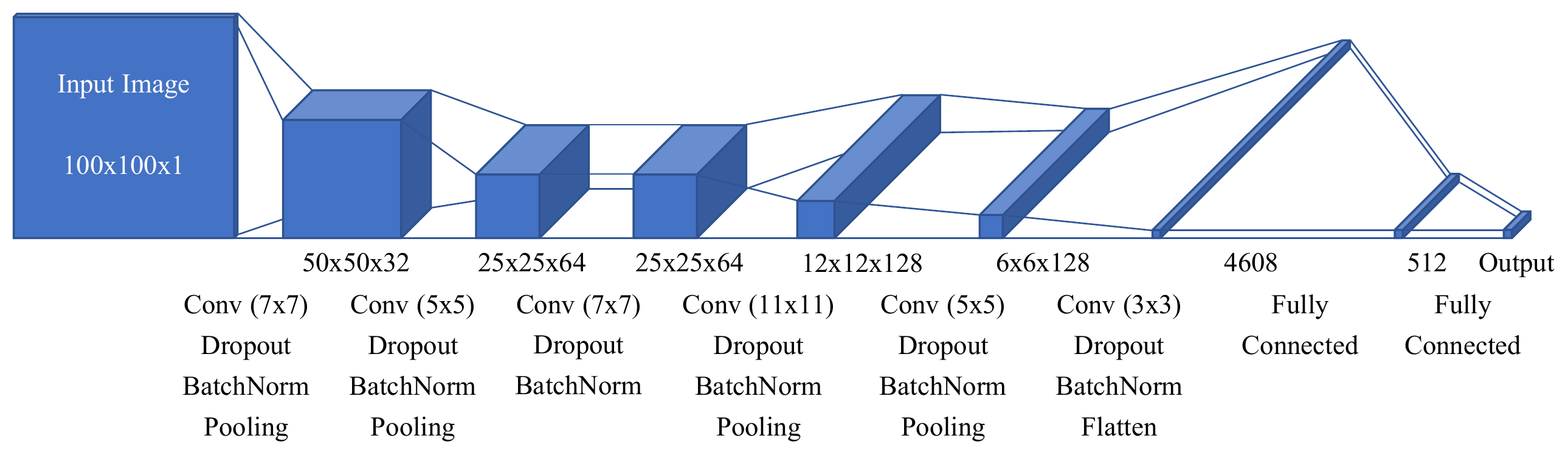}
    \caption{Structure of the CNN used in this work, showing the input image and the output of each block of layers. In total there are six convolutional layers, each with dropout and batch normalisation, four max-pooling layers and two fully connected layers. A 'flatten' layer is also included to connect the multidimensional data to the 1D fully connected layer, and ReLU activation is used throughout. Numbers given above or beside each output are the output shapes. The types of layers in each block are given underneath, along with the kernel sizes (in pixels) used. Further details can be found at the end of Section \ref{subsec:cnn_architecture}.}
    \label{fig:cnn_structure}
\end{figure*}

For mass model parameter estimation, the network used in this work contains over 4 million trainable weights and biases, featuring six convolutional layers, six dropout layers, four pooling layers and two fully connected layers, as shown in Fig. \ref{fig:cnn_structure}. Convolutional layers convolve the input with a series of kernels, as described above, with the biases and weights of each kernel optimised through training. These layers have manually-tuneable hyperparameters such as the number and size of the kernels, and through experimentation it was found that six convolutional layers with the kernel sizes given in the figure provided the best efficiency (in terms of accuracy and training time) for this work. We chose a stride of one and incorporated zero-padding to maintain the image size through the convolution. The activation function used throughout is the Rectified Linear Unit \citep[ReLU;][]{nair2010rectified}, which acts non-linearly on the nodes such that any negative values are set to zero.

Dropout \citep{srivastava2014dropout} is applied after each convolutional layer (see Section \ref{subsec:cnn_uncertainties}), after which batch normalisation is performed, which normalises the output of these layers to increase the stability of the network. The pooling layers used in this work are max-pooling, which output new arrays containing the maximum values of each two-by-two-sized region of the input (with a stride of two), and are hence used to decrease the size of the input. This allows later network layers to examine larger, more abstract features while also reducing computation time. There is also a 'flatten' layer used in this network that converts the three-dimensional data into a one-dimensional vector. This is passed to the fully connected layers, whose nodes are each connected to every node in the preceding layer, which are used to identify relationships in the data that the convolutional layers cannot.

The final layer of the network is the output layer. The network predicts values for the parameters of a lens mass model fitted to the lensed image; for an SIE lens the predicted parameters are the Einstein radius and the two components of complex ellipticity, given as
\begin{equation}
    e_{1}=\frac{1-q^{2}}{1+q^{2}}\cos{2\phi}\,, \\ e_{2}=\frac{1-q^{2}}{1+q^{2}}\sin{2\phi}\,,
    \label{eq:complex_ellip}
\end{equation}
where $\phi$ and $q$ are the orientation and axis ratio of the mass profile, respectively. When fitting the lens with a power law profile, the slope parameter $n$ is also predicted.
To improve the network's performance, the inputs are pre-processed; the input images are all intensity scaled so the counts in each pixel lie in the range 0-1, and the training parameters to be predicted are all rescaled to lie in the same ranges as each other, between zero and ten. The parameters predicted by the CNN are then rescaled back again for calculating accuracy values, with complex ellipticity converted to the orientation and axis ratio of the lens mass profile.

To evaluate the performance of the network during training, a cost function is required that gives a measure of the difference between the CNN's predictions and the true values. This is often taken to be the mean squared error of the predicted mass model parameters compared to their true values.
However, the inclusion of an approximate Bayesian framework for uncertainty predictions necessitates that another form of cost function be used throughout this work (see Section \ref{subsec:cnn_uncertainties}).

\subsubsection{Uncertainties}
\label{subsec:cnn_uncertainties}

Standard CNNs for regression problems can be trained to predict values, yet there is no way of reliably obtaining uncertainties on such values. However, work by \citep{gal2016bayesian,gal2016dropout,kendall2017uncertainties} has seen the incorporation of Bayesian statistics into neural networks, in which a trained network's weights can vary according to a probability distribution rather than remaining fixed, allowing for such a network to obtain posterior probabilities and hence overall uncertainties. \cite{levasseur2017uncertainties} have since applied this to create an approximate Bayesian CNN framework for strong lens modelling, involving the use of variational inference. For completeness, we include here the key information and equations, and leave the bulk of the theory to that paper.

In variational inference, a variational distribution $q(w)$ over a set of unobserved weights $w$ is used to approximate the posterior $p(w|X,Y)$ of the weights given training data $(X,Y)$. In \cite{levasseur2017uncertainties}, they choose a form of $q(w)$ such that sampling from it is equivalent to performing dropout over network weights.
Introduced by \cite{srivastava2014dropout}, dropout is the process of randomly "dropping" nodes (setting them to zero) temporarily for each forward pass through the network during training. This stops the network from becoming too reliant on certain pathways in order to prevent it from overfitting to its training data.
After training, performing inference using the approximated posterior predictive distribution, $p(y|x,X,Y)=p(y|x,w)q(w)dw$ for a set of test data $(x,y)$ can be done by approximating this integral with a Monte Carlo (MC) integral, through repeatedly passing each test image through the network and using dropout during {\it testing} (MC dropout) to sample from the approximate parameter posterior. This sampling thus gives a measure of the \textit{epistemic} uncertainty (how well the network has been trained).

The other type of uncertainty to be obtained is the \textit{aleatoric} uncertainty, which describes errors coming from the input data itself, such as the noise level of images.
This is achieved through the modification of the network's cost function, choosing its form to be the negative of a Gaussian log-likelihood
\begin{equation}
    -\mathcal{L}=\sum_{k}\frac{1}{2}\left\Vert{y_{{\rm pred}, k}-y_{{\rm true}, k}}\right\Vert^2\exp{(-s_k)}+\frac{1}{2}s_k
    \label{eq:cost_function}
\end{equation}
where $y_{{\rm pred}, k}$ and $y_{{\rm true}, k}$ are the predicted and true values of parameter $k$ for a given training image, and $s_k=\log\sigma_k^2$ is the log-variance.
$\sigma_k$ represents the aleatoric uncertainty of each parameter, and due to its presence in both terms it is optimised through training and can be outputted for each image fed through the network. We train using the log-variance rather than $\sigma_k$ itself in order to avoid potential division by zero and to improve numerical stability. Through this process, the CNN can now be trained to output both predictions of the parameter values and predictions of their associated aleatoric uncertainties.
Using the MC dropout method, these are then added in quadrature with epistemic uncertainties to give the overall measure of the CNN's uncertainty for a given image.

\subsubsection{Training}
\label{subsec:cnn_training}

\begin{figure*}
    \begin{tabular}{cc}
    \includegraphics[width=\columnwidth]{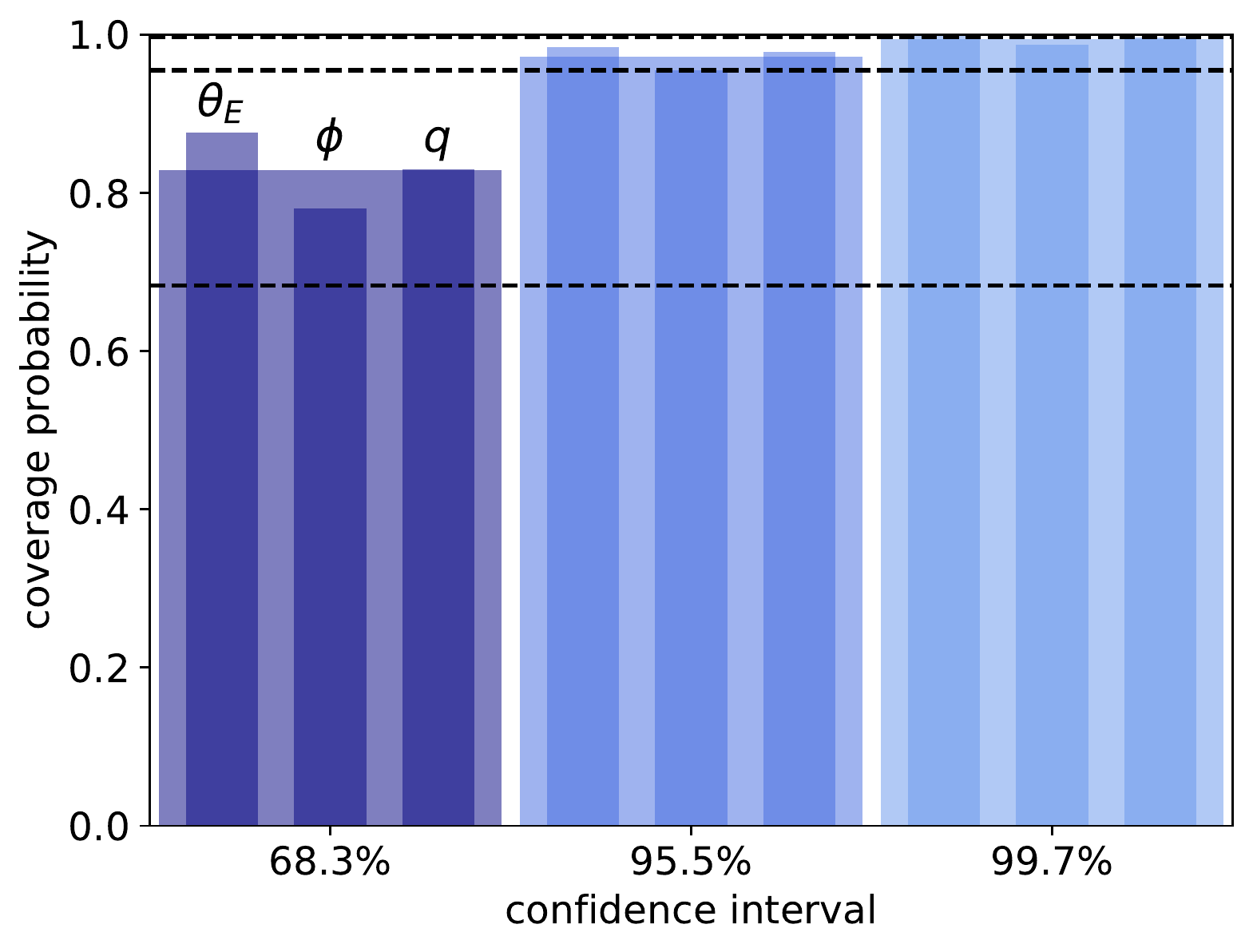} & \includegraphics[width=\columnwidth]{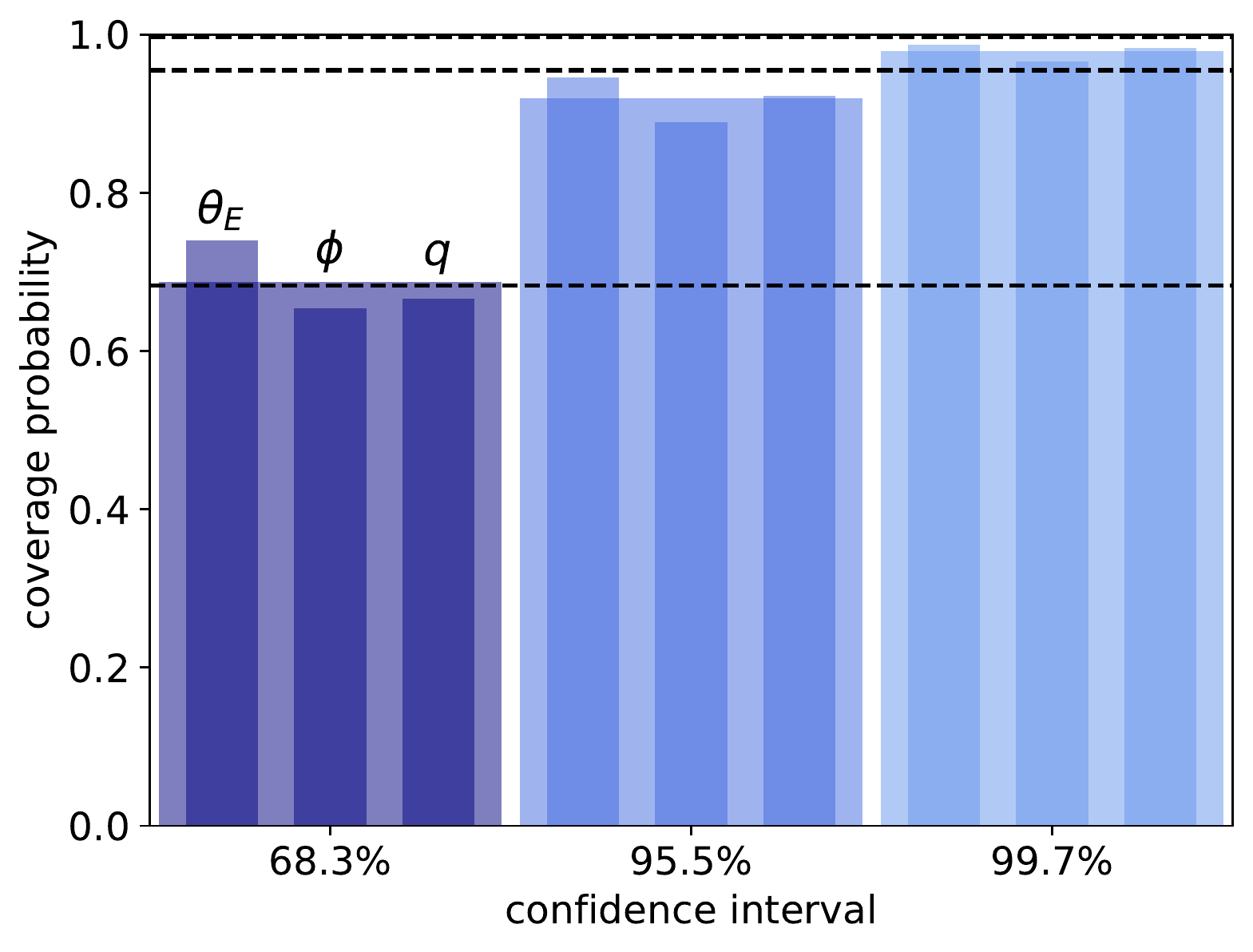} \\
    \includegraphics[width=\columnwidth]{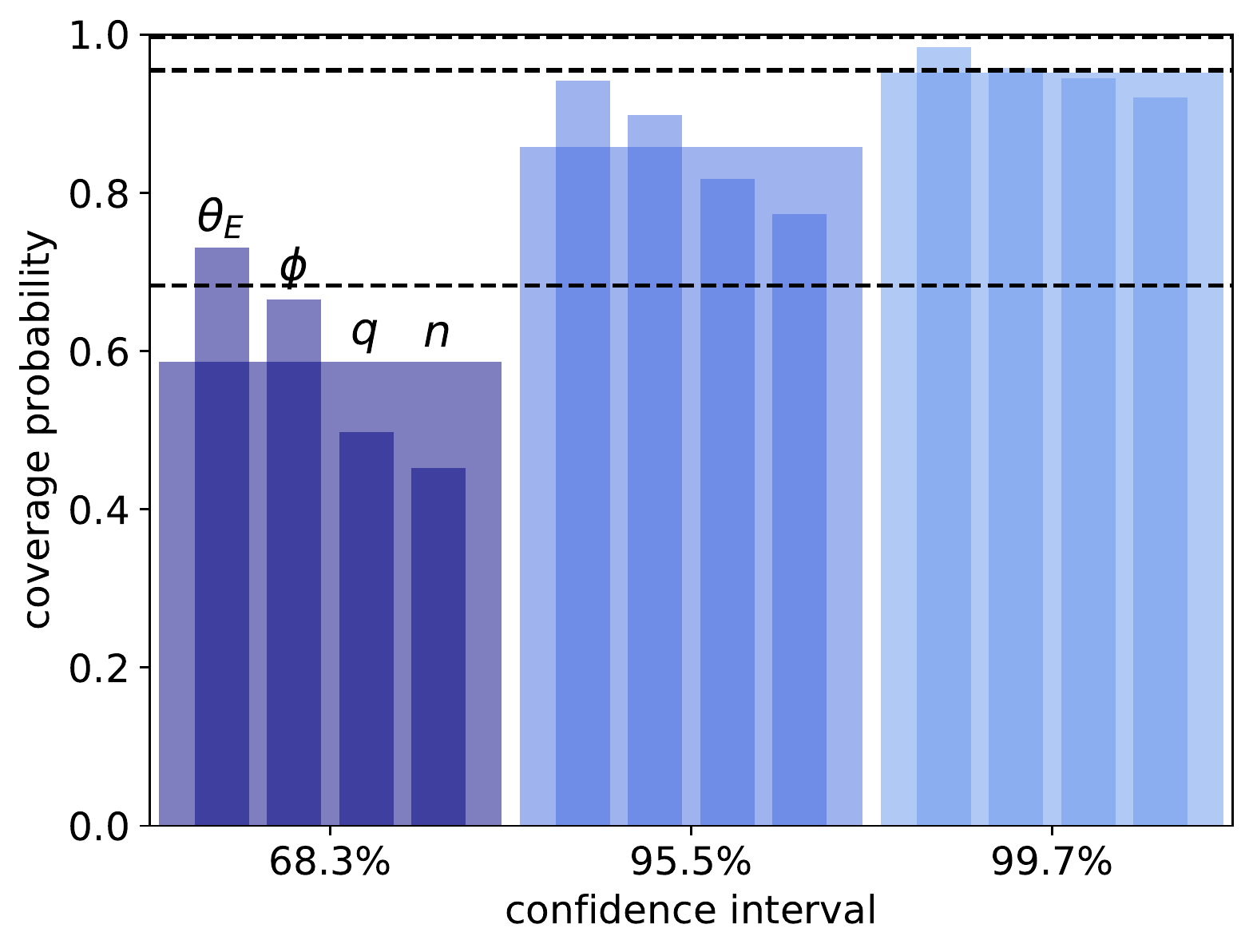} & \includegraphics[width=\columnwidth]{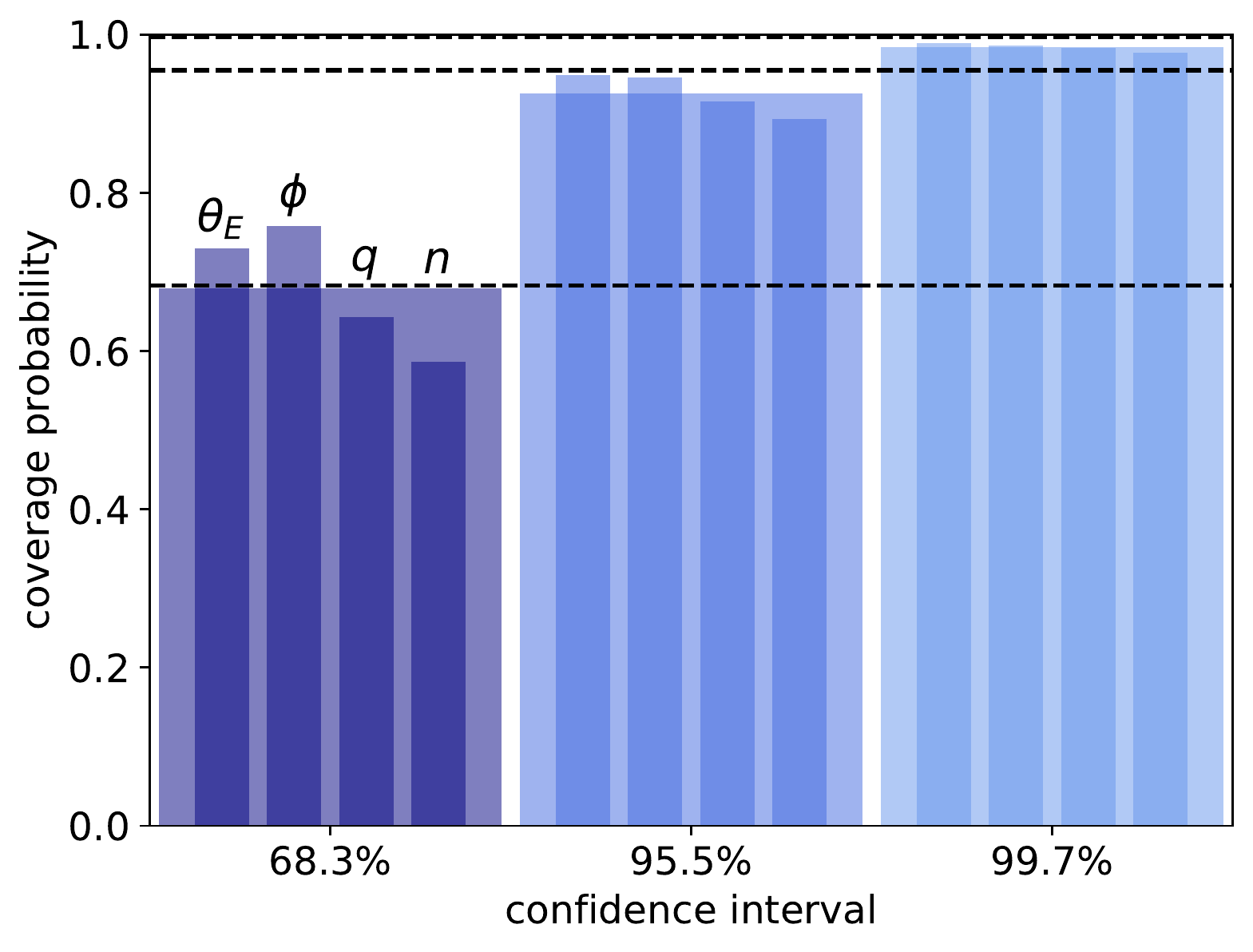} \\
    \end{tabular}
    \caption{Coverage probabilities when fine-tuning dropout during CNN training to ensure appropriately-sized predicted uncertainties.
    Top and bottom rows show results for the CNN trained to predict SIE parameters ($\theta_{E}$, $\phi$, $q$) and power law parameters ($\theta_{E}$, $\phi$, $q$, $n$), respectively, indicated by the smaller bars.
    Average coverage probabilities are given as the wider bars, and for the ideal case would reach each dashed line representing 68.3, 95.5 and 99.7 per cent coverage.
    Left: example results from using incorrect dropout rates. Right: results from using fine-tuned dropout rates.}
    \label{fig:coverage_probs}
\end{figure*}

To train the network, 100,000 images were fed into the network in batches of 100, along with a validation set of 10,000 images. These images were 100$\times$100-pixel "postage stamp" cutouts of lenses, and featured a range of source types in order for the network to cope with the wide range of realistic sources (see Section \ref{subsec:training_sims}).
Two CNNs were trained: one to predict parameters for SIE lenses and a second for power law lenses. They featured the same architecture (with one extra output for the power law slope) and used the negative log-likelihood cost function from Section \ref{subsec:cnn_uncertainties}. For network optimization, the best results were obtained using a learning rate of 1e-3 with the Nadam optimizer \citep{dozat2015incorporating}, which is a combination of methods based on the stochastic gradient descent algorithm.

The CNN was run on a GPU machine, allowing for much faster processing; training over 150 epochs on 100,000 images took less than two hours, while testing on thousands of images took on the order of a few seconds to minutes depending on the number of repeats performed to obtain uncertainties from dropout during testing. For the results presented in this work each image was tested 100 times to give a good level of accuracy in a manageable amount of time.

In order for the CNN to predict suitable uncertainties, the dropout rate must be fine-tuned. After training with a given dropout rate, the CNN was tested on 10,000 images covering parameter space with the same imaging characteristics as the training set. Coverage probabilities were obtained for each parameter; these give the fraction of the test set whose predicted values lie within a given confidence interval of the true value. These confidence intervals are determined based on the predicted 1$\sigma$ uncertainties, also scaled up to 2$\sigma$ and 3$\sigma$ intervals. Hence, the ideal case would consist of 68.3 per cent of predictions lying within the 1$\sigma$ confidence interval, and so on. Fine-tuning was achieved through repeated training and testing with different dropout rates in order to achieve the above, with a final dropout rate of 5 per cent (i.e. a keep rate of 95 per cent).
Some example coverage probabilities are given in Fig. \ref{fig:coverage_probs}.

\subsection{\textsc{PyAutoLens} + CNN}
\label{subsec:pyal_plus_cnn}

The parametric fitting of \textsc{PyAutoLens} can model the lens and source to a high degree of accuracy, but to do so requires a large amount of time and computing power as well as the manual effort of setting priors on the parameters. These are factors we would like to reduce, as they currently limit its applicability to the quantity of strong galaxy-galaxy lenses that future surveys will discover.
\textsc{PyAutoLens} can be crudely automated by setting priors to cover parameter space rather than basing them on human inspection ('blind' modelling), however this results in a drop in accuracy and further increased modelling time. In addition, the widths of the priors can present problems - if the bounds are too narrow, the software may not converge to the correct solution, while too large bounds result in many more iterations before a solution is obtained. Even then, wide bounds on the priors can result in the software converging on parameters that represent overmagnified or undermagnified solutions rather than recovering the true parameters \citep{maresca2020auto}.

On the other hand, a trained CNN provides a much faster, computationally inexpensive and automated alternative, but requires a sufficiently large and realistic training set in order to avoid underfitting or overfitting to the data. Given this, the CNN may not always be able to reach the high accuracy that \textsc{PyAutoLens} can achieve with human inspection, but for the automated blind modelling required by \textsc{PyAutoLens} to cope with future surveys the CNN can provide a more efficient and accurate method.
Additionally, CNN predictions can act as priors for \textsc{PyAutoLens}, with such a combination of methods removing the need for human inspection and allowing for slower, yet still automated, accurate modelling to deal with more complex lenses.

\textsc{PyAutoLens} initialises lens model parameters by randomly sampling from a series of user-defined prior distributions which can be uniform or Gaussian. When setting up, these are typically manually selected based on a visual inspection of the lensed image. Automation of this process would require either generous uniform priors to ensure that no parameters are disallowed during fitting, or a means of making an estimation of a narrower set of priors.
Since the CNN can now predict values for mass model parameters extremely quickly, this opens up the possibility of using these to inform the priors for \textsc{PyAutoLens}. Not only this, but the uncertainties that the CNN predicts can be used as the 1$\sigma$ bounds, now using a Gaussian prior distribution (centred on the CNN's predicted parameters) rather than a uniform one. 

Such a combination of CNNs and conventional parametric modelling can help alleviate the shortfalls that each method faces individually: CNN accuracy is limited by the quality of the training set, which ideally would feature non-parametric source and lens profiles for many hundreds of thousands of images, and without these the network may struggle with more complex lensing systems. 
Meanwhile, conventional parametric modelling does not require training, and as such offers a more flexible means of modifying the lensing model, e.g., changing the mass profile and the easy addition of components such as external shear. However, this modelling can take a large number of iterations before converging on a solution, and may require repeat modelling attempts.

Together, CNNs can vastly simplify the search over parameter space for parametric modelling methods and can additionally prevent methods using semilinear inversion, like \textsc{PyAutoLens}, from falling into local minima, especially when using CNN uncertainties to reduce the size of prior parameter space. Hence this combination may lead to a fully automated pipeline that can obtain accurate results over short time-scales for application to large samples of lenses.

\section{Results}
\label{sec:results}

In this section, the performance of the CNN is compared to that of \textsc{PyAutoLens} modelling blindly (PyAL (blind)) for a range of test cases with increasing complexity.
In addition, both are compared to combinations of the two techniques, in which the CNN predictions are used as priors for \textsc{PyAutoLens}. The first (PyAL + CNN) uses uniform priors of arbitrary width that are centred on the CNN-predicted model parameters.
The second (PyAL + CNN (1$\sigma$)) uses Gaussian priors centred on these parameters with the CNN-predicted 1-sigma uncertainties acting as the prior widths; see Section \ref{subsec:pyal_plus_cnn}.
We also tried setting the prior widths to twice the 1-sigma uncertainties reported by the CNN in case the uncertainties were underpredicted for the more complex test sets, but overall the results were worse than passing the unmodified uncertainties and as such are not included in this paper.
The uniform prior distributions of PyAL (blind) covered parameter space for all parameters except Einstein radius, which instead used $\pm0.2$ arcsecond bounds centred on the approximated Einstein radii; see Section \ref{subsec:pyautolens}.
For the PyAL + CNN method, the bounds were $\pm0.2$ arcseconds, $\pm40^{\circ}$, $\pm0.1$, and $\pm0.2$ for Einstein radius, orientation, axis ratio and slope, respectively.

In Sections \ref{subsec:sie_parametric} and \ref{subsec:sie_hudf}, the modelling techniques are applied to simulated images containing SIE lenses, with parametric (S\'ersic and Gaussian) and real HUDF sources, respectively. In the case of HUDF sources, results are compared for lenses with and without LOS structure, and this is repeated for the case of power law lenses in Section \ref{subsec:pwrlaw_hudf}. In Section \ref{subsec:eagle_hudf}, the techniques are instead applied to simulated images with complex lenses in the form of galaxies obtained from the EAGLE hydrodynamical simulation, again with HUDF sources and for those with and without LOS structure.

Throughout the remainder of this paper, we quantify the error on predicted lens model parameters by the 68 per cent confidence interval computed from the distributions of differences between true and predicted parameter values across the test image set. In our previous work we referred to this 68 per cent confidence interval as a parameter's 'uncertainty', however in this work we hereafter refer to this as a parameter's 'error' to avoid confusion with the CNN's predicted 1$\sigma$ uncertainties.

\subsection{SIE lenses + parametric sources}
\label{subsec:sie_parametric}

\begin{figure*}
  \centering
  \begin{minipage}[t]{\columnwidth}
    \includegraphics[scale=0.44]{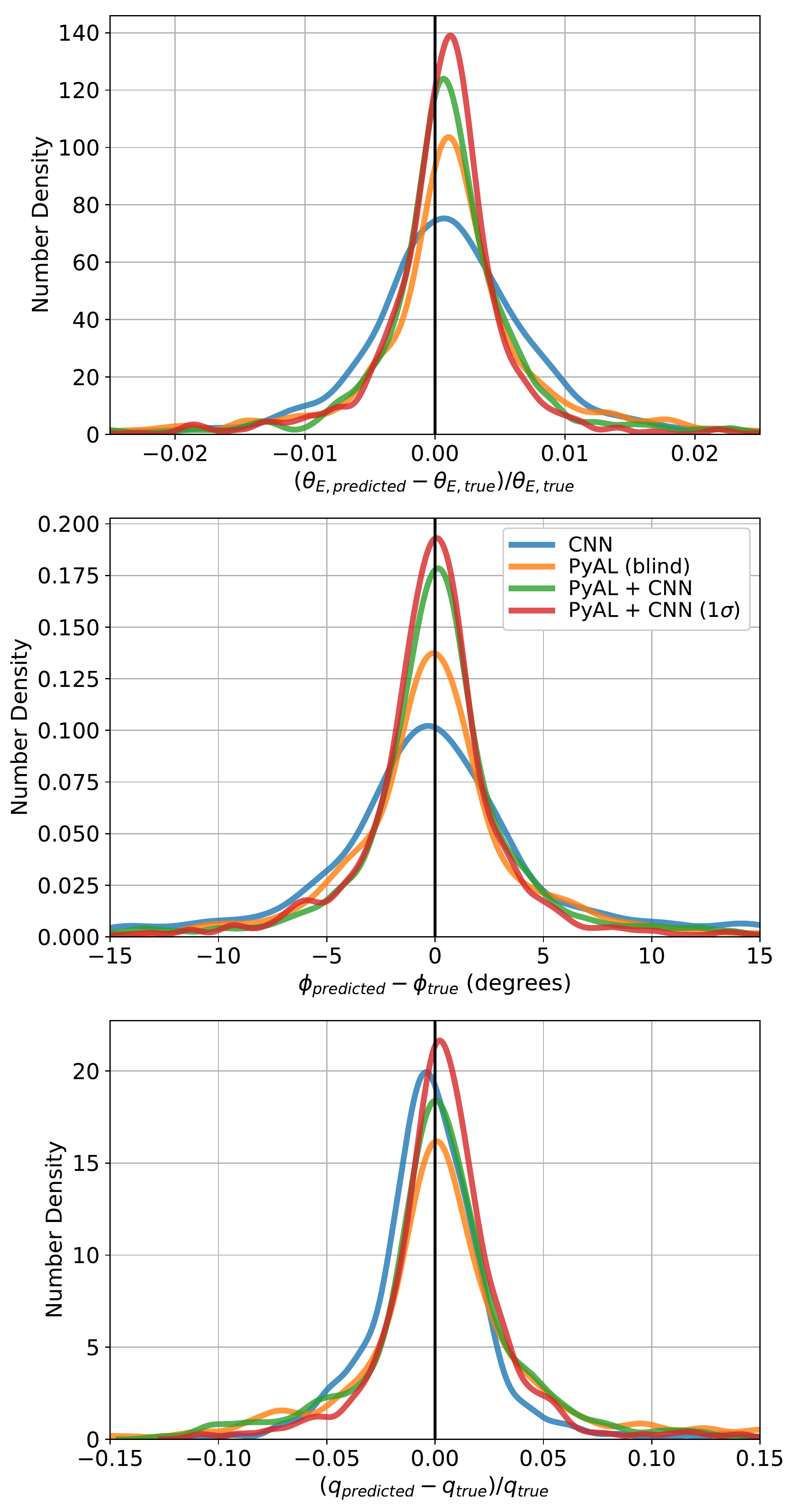}
    \caption{Distribution of the differences between predicted SIE mass model parameters and their true values for a test data set of 1000 images containing parametric sources: 500 with S\'ersic sources and 500 with Gaussian sources. From top to bottom: Einstein radius, orientation and axis ratio of the lens mass profile. Einstein radius and axis ratio results presented as fractional differences. The distributions shown are those for the CNN (blue), \textsc{PyAutoLens} modelling blindly (orange), \textsc{PyAutoLens} using CNN predictions as priors (green) and \textsc{PyAutoLens} using CNN predictions and 1-sigma uncertainties as priors (red).}
    \label{fig:kde_sie_parametric}
  \end{minipage}
  \hfill
  \begin{minipage}[t]{\columnwidth}
    \includegraphics[scale=0.44]{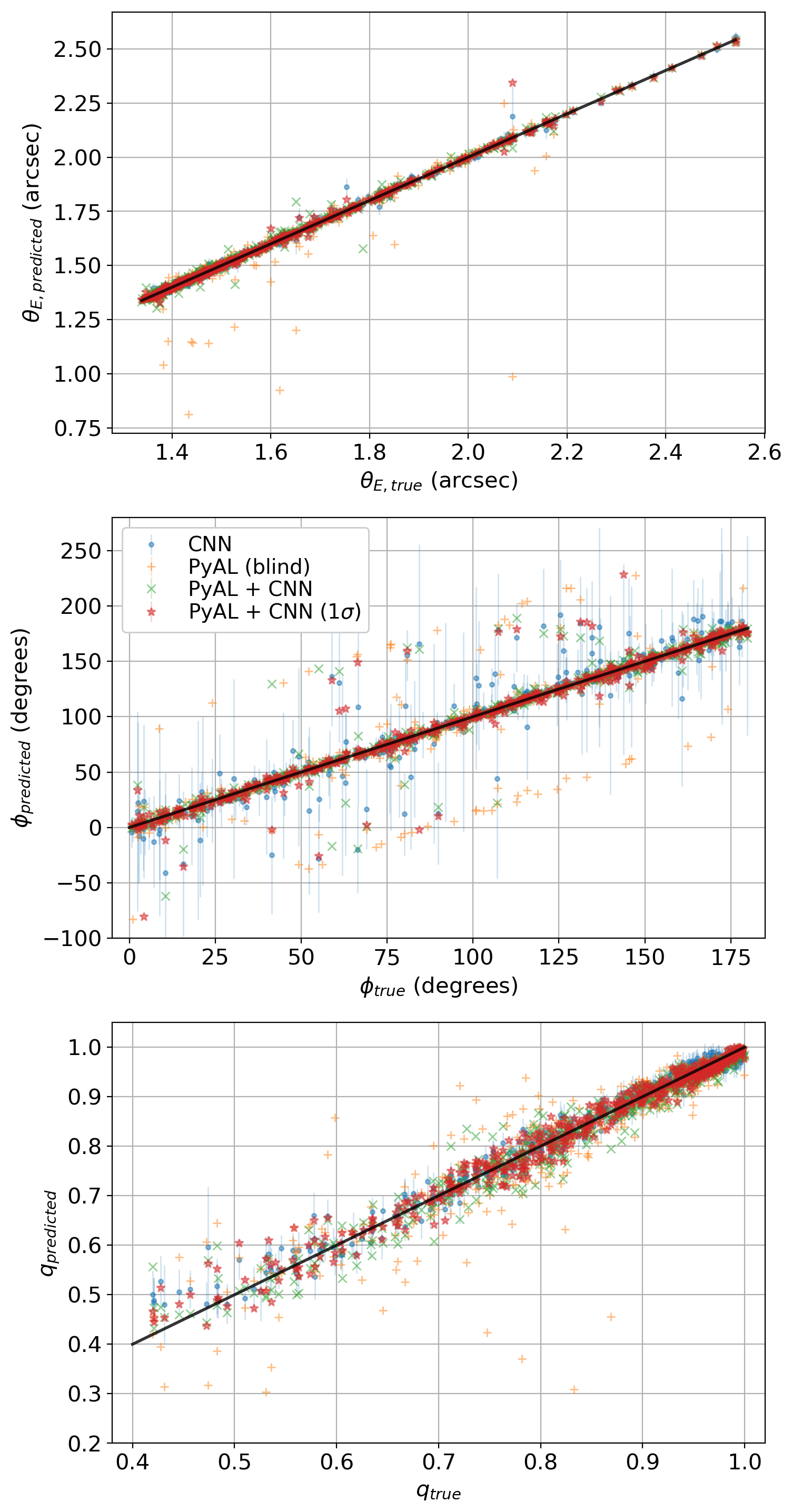}
    \caption{Comparison of predicted SIE lens parameters with the true values for test data sets of 1000 images containing SIE lenses and parametric (S\'ersic and Gaussian) sources. From top to bottom: Einstein radius, orientation and axis ratio of the lens mass profile.}
    \label{fig:predvstrue_sie_parametric}
  \end{minipage}
\end{figure*}

We began by testing for any inherent biases or differences between the modelling methods, having them model 1000 images with SIE lenses and parametric sources. Half contained S\'ersic sources while the other contained Gaussian sources.
Results are presented in Fig. \ref{fig:kde_sie_parametric} as distributions of the differences between predicted values and true values for each mass model parameter. Fractional differences are used for Einstein radius and axis ratio. Table \ref{tab:sie_parametric} gives the 68 per cent confidence intervals of these distributions. Scatter plots of predictions against true values are given in Fig. \ref{fig:predvstrue_sie_parametric}, including the CNN's predicted 1$\sigma$ uncertainties. 
It is clear from the results that the CNN's predicted uncertainties are of an appropriate size and accurately reflect its errors; the large uncertainties in some orientation values are expected as they occur for round lenses with no well defined orientation.
It is worth noting that the CNN errors are significantly lower than those presented in \cite{pearson2019use}; this is primarily due to doubling the size of the training set, which now also contains a greater variety of lensing systems. The incorporation of Dropout and uncertainty prediction into the network architecture, along with testing on lenses with generally larger Einstein radii (and hence more available information), also serve to increase network accuracy.

\begin{table}
	\centering
	\caption{\textbf{SIE lenses + parametric sources.} The 68 per cent confidence intervals on predicted parameters for each modelling method, computed from the distributions of differences between true and predicted parameter values across 1000 test images.}
	\label{tab:sie_parametric}
	\begin{tabular}{llll}
	    \hline
        Method & $\theta_E$ (arcsec) & $\phi$ ($^{\circ}$) & $q$\\
		\hline
        CNN & 0.0090 & 4.92 & 0.018 \\
		PyAL (blind) & 0.0089 & 4.44 & 0.027 \\
		PyAL + CNN & 0.0065 & 2.81 & 0.021 \\
		PyAL + CNN (1$\sigma$) & 0.0055 & 2.47 & 0.017 \\
		\hline
	\end{tabular}
\end{table}

The CNN by itself achieves a higher accuracy than that of \textsc{PyAutoLens} for axis ratio, but lower accuracies for the other parameters. Despite the significantly lower peaks for Einstein radius and orientation the CNN has far fewer outliers, as shown in Fig. \ref{fig:predvstrue_sie_parametric}, resulting in similar overall errors. However, for all mass model parameters the combination of the two methods does significantly better than \textsc{PyAutoLens} modelling blindly, with PyAL + CNN (1$\sigma$) achieving 36-44 per cent lower errors than PyAL (blind). There is also a notable benefit to including the CNN's predicted 1$\sigma$ uncertainties in the priors of \textsc{PyAutoLens}, with 12-17 per cent lower errors for PyAL + CNN (1$\sigma$) compared to PyAL + CNN.
While there are biases observed in the distributions towards underpredicting or overpredicting parameters, the magnitudes of these biases are of the order <0.2 per cent for Einstein radius and <1 per cent for axis ratio, and so have negligible impact on modelling.

\subsection{SIE lenses + HUDF sources}
\label{subsec:sie_hudf}

\begin{figure*}
  \centering
  \begin{minipage}[t]{\columnwidth}
    \includegraphics[scale=0.44]{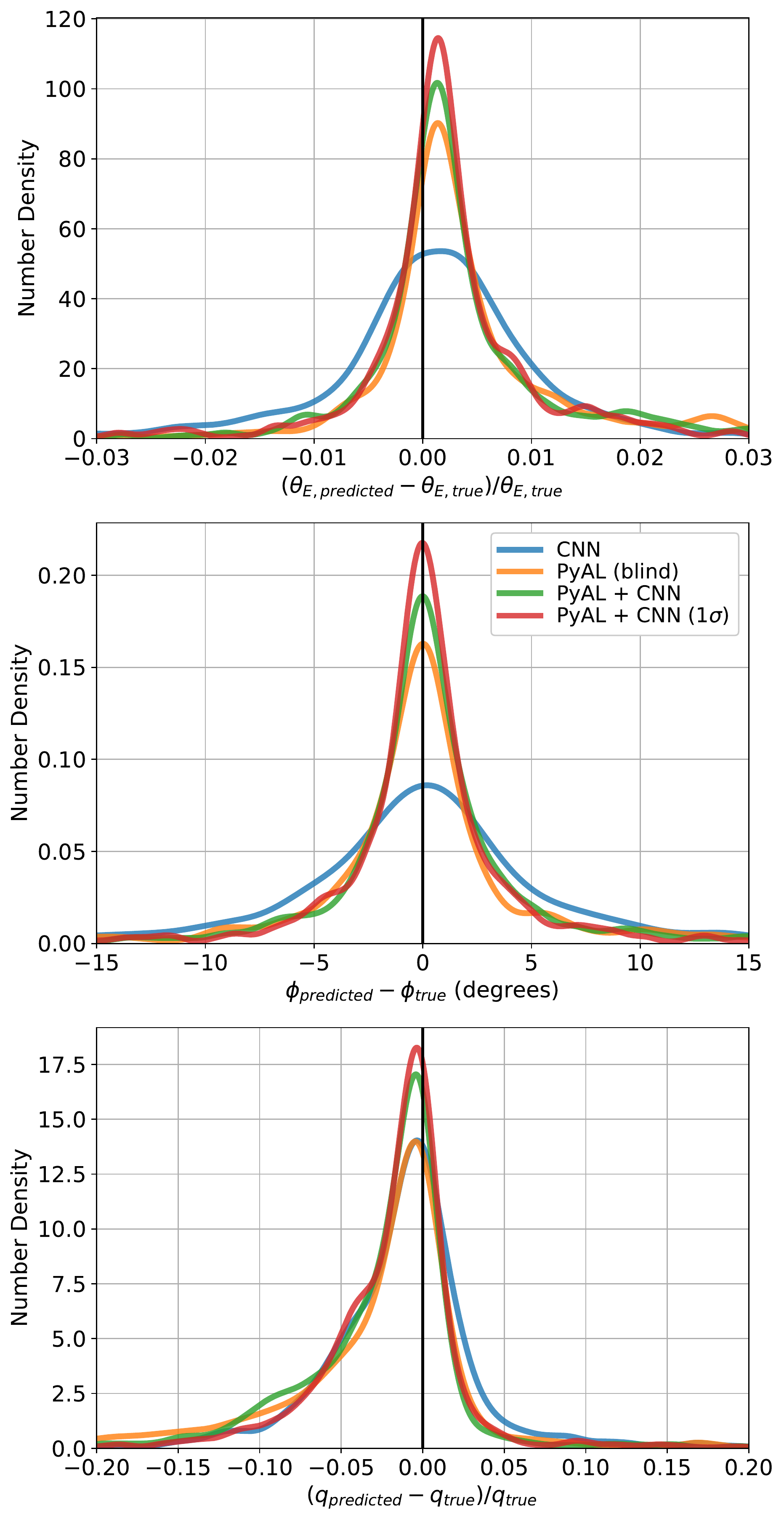}
    \caption{Distribution of the differences between predicted SIE mass model parameters and their true values for test data sets of 1000 images containing HUDF sources, without LOS structure. 
    The distributions follow the same format as Fig. \ref{fig:kde_sie_parametric}.}
    \label{fig:kde_sie_hudf_nolos}
  \end{minipage}
  \hfill
  \begin{minipage}[t]{\columnwidth}
    \includegraphics[scale=0.44]{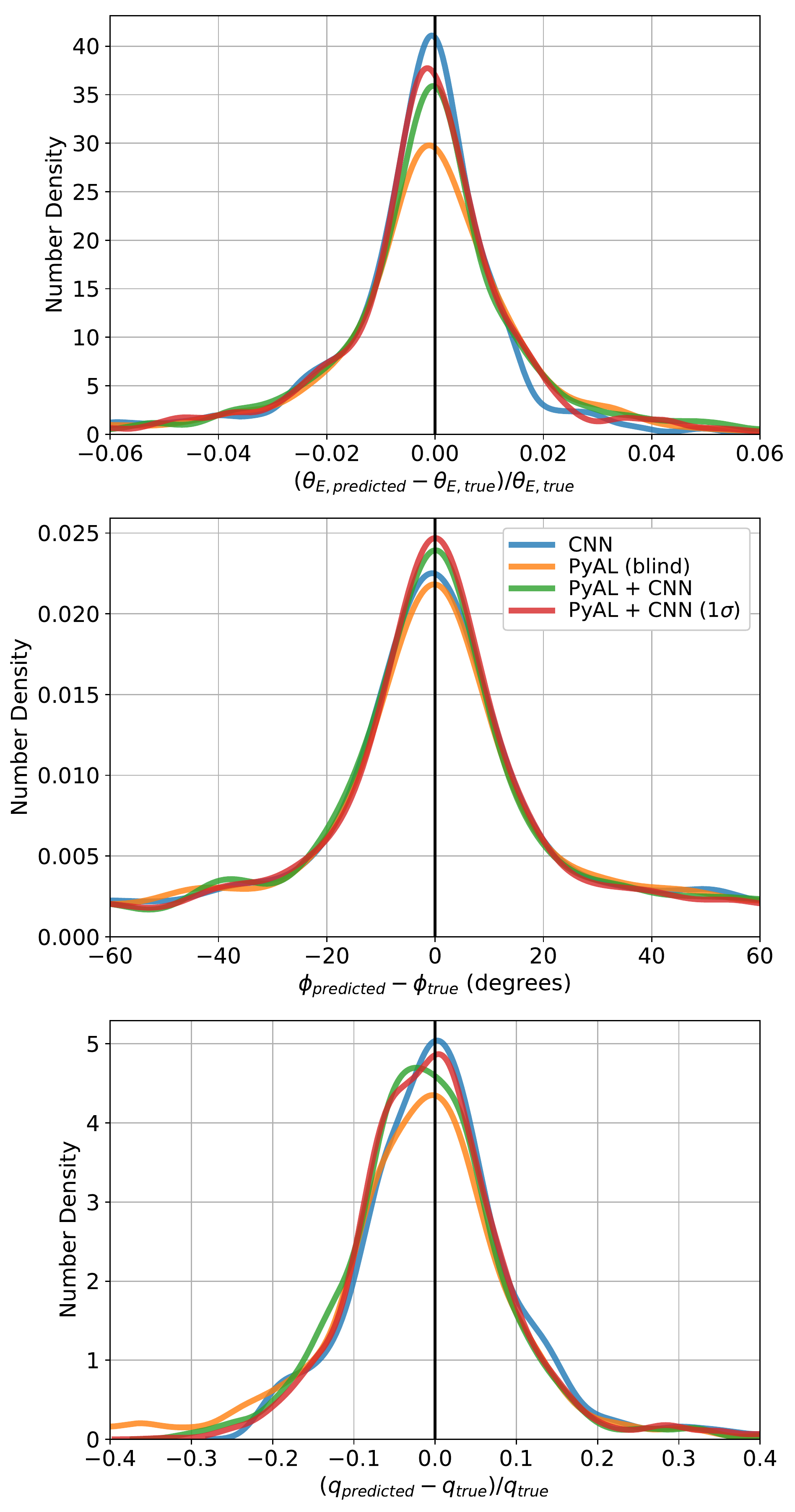}
    \caption{Distribution of the differences between predicted SIE mass model parameters and their true values for test data sets of 1000 images containing HUDF sources, with LOS structure.
    The distributions follow the same format as Fig. \ref{fig:kde_sie_parametric}.}
    \label{fig:kde_sie_hudf_los}
  \end{minipage}
\end{figure*}

\begin{figure*}
  \centering
  \begin{minipage}[t]{\columnwidth}
    \includegraphics[scale=0.44]{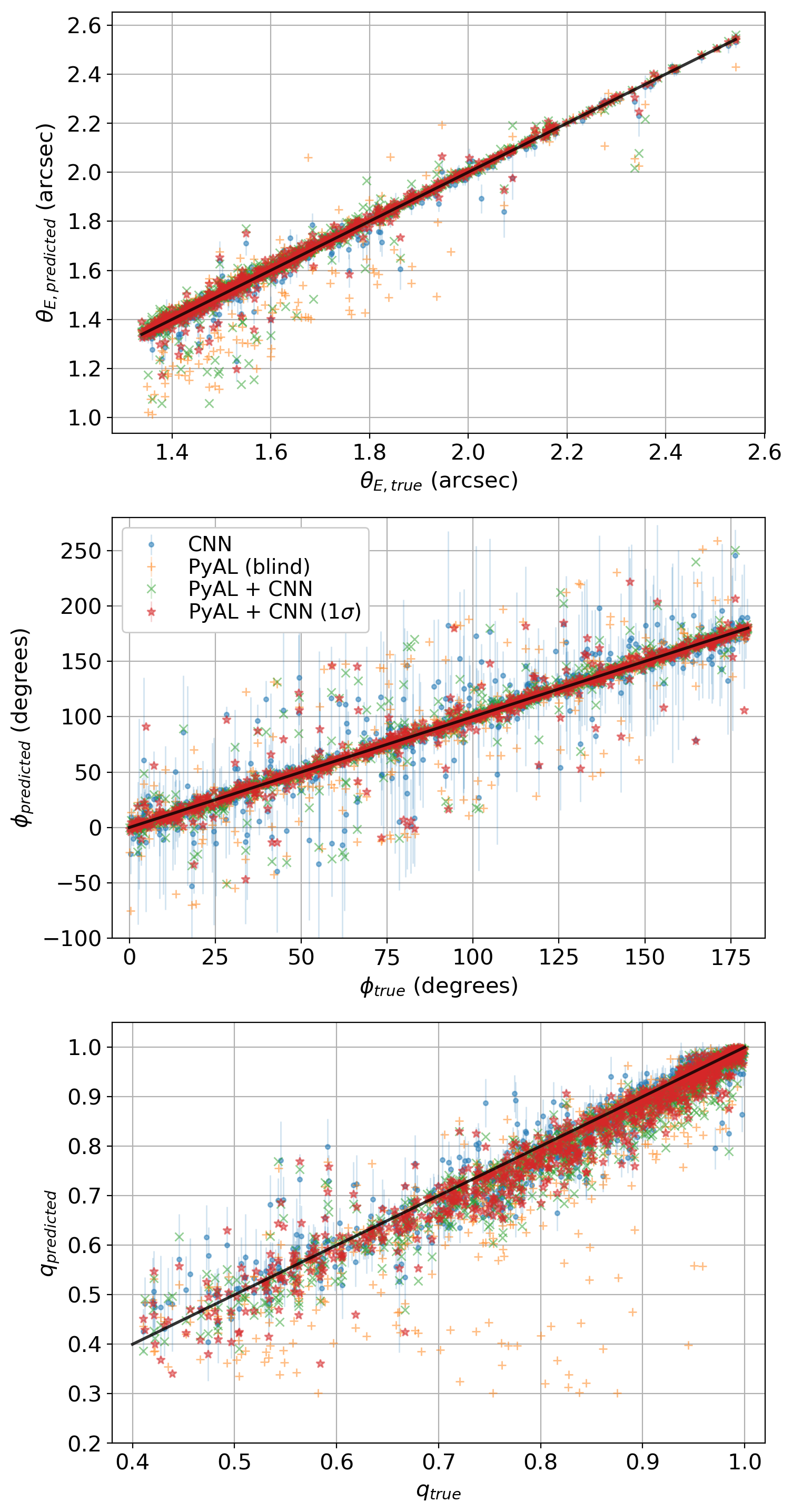}
    \caption{Comparison of predicted SIE lens parameters with the true values for test data sets of 1000 images containing SIE lenses and HUDF sources, without LOS structure. From top to bottom: Einstein radius, orientation and axis ratio of the lens mass profile.}
    \label{fig:predvstrue_sie_hudf_nolos}
  \end{minipage}
  \hfill
  \begin{minipage}[t]{\columnwidth}
    \includegraphics[scale=0.44]{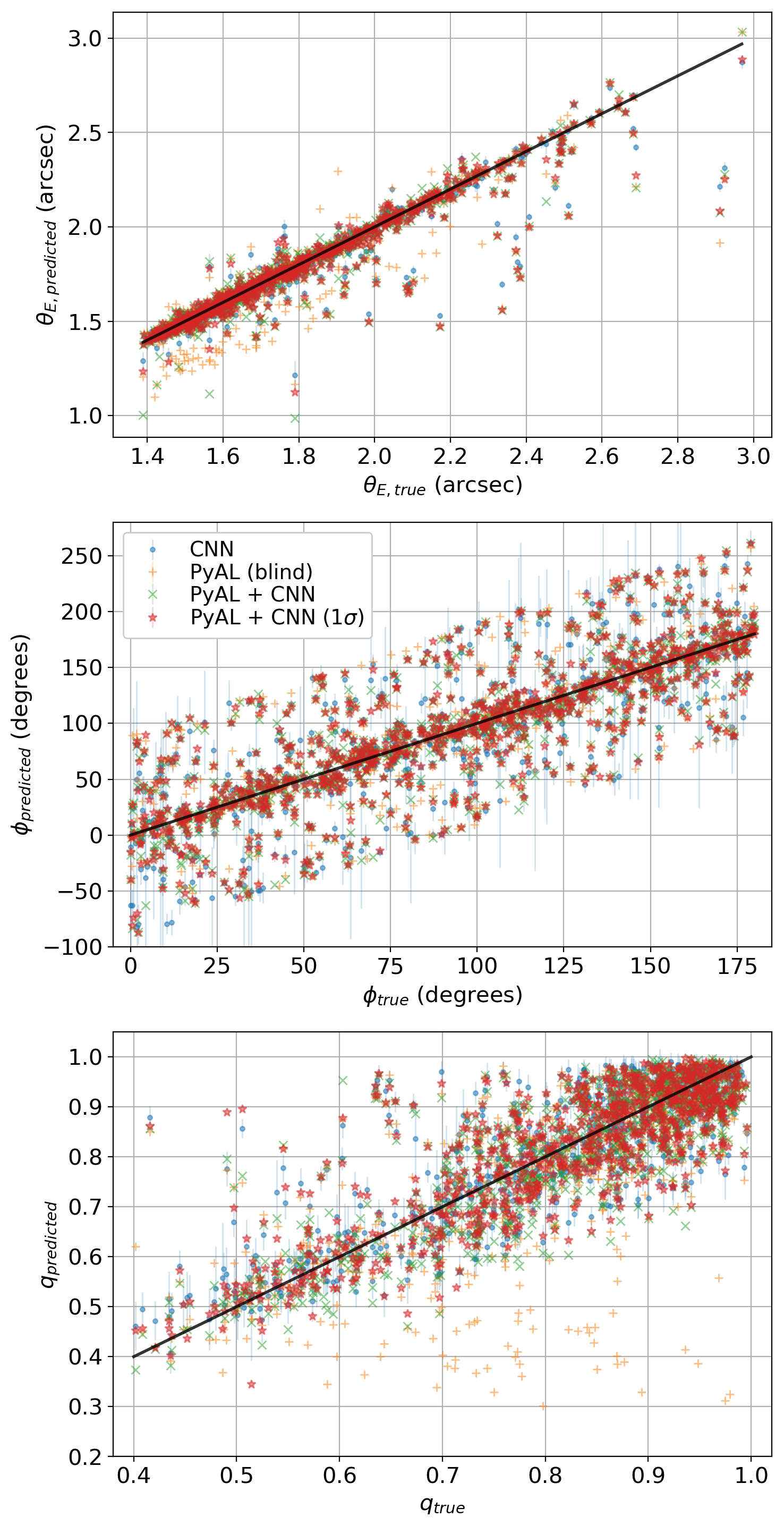}
    \caption{Comparison of predicted SIE lens parameters with the true values for test data sets of 1000 images containing SIE lenses and HUDF sources, with LOS structure. From top to bottom: Einstein radius, orientation and axis ratio of the lens mass profile.}
    \label{fig:predvstrue_sie_hudf_los}
  \end{minipage}
\end{figure*}

\begin{table}
	\centering
	\caption{\textbf{SIE lenses + HUDF sources.} The 68 per cent confidence intervals on predicted parameters for each modelling method, computed from the distributions of differences between true and predicted parameter values across 1000 test images.}
	\label{tab:sie_hudf}
	\begin{tabular}{llll}
        \hline
        Method & $\theta_E$ (arcsec) & $\phi$ ($^{\circ}$) & $q$\\
        \hline
		{\bf Without LOS structure} & & & \\
        CNN & 0.013 & 6.10 & 0.029 \\
		PyAL (blind) & 0.013 & 4.22 & 0.038 \\
		PyAL + CNN & 0.010 & 3.02 & 0.030 \\
		PyAL + CNN (1$\sigma$) & 0.009 & 2.57 & 0.026 \\
		{\bf With LOS structure} & & & \\
        CNN & 0.021 & 30.7 & 0.067 \\
		PyAL (blind) & 0.029 & 33.8 & 0.077 \\
		PyAL + CNN & 0.026 & 28.0 & 0.068 \\
		PyAL + CNN (1$\sigma$) & 0.023 & 27.1 & 0.065 \\
		\hline
	\end{tabular}
\end{table}

The next set of testing involved examining the impact of both using real HUDF sources in place of simple parametric sources, and incorporating additional structure along the LOS. Results for images without LOS structure are presented in Figs \ref{fig:kde_sie_hudf_nolos} and \ref{fig:predvstrue_sie_hudf_nolos}, and those with LOS structure in Figs \ref{fig:kde_sie_hudf_los} and \ref{fig:predvstrue_sie_hudf_los}, following the same format as before.
68 per cent confidence intervals are given in Table \ref{tab:sie_hudf}.

We first examine the results for images without LOS structure.
In general, the accuracies for all modelling have dropped compared to testing on images with parametric sources, as expected. As with the previous test set containing parametric sources, the CNN accuracy is lower for Einstein radius and orientation and higher for axis ratio when compared to \textsc{PyAutoLens} blind modelling. Likewise, we again see the combination of the two techniques performing significantly better than either of them separately, with PyAL + CNN (1$\sigma$) errors 31-39 per cent lower than PyAL (blind).
In Fig. \ref{fig:predvstrue_sie_hudf_nolos}, there are a notable number of axis ratio values that PyAL (blind) significantly underpredicts that the other methods do not. These scattered outliers generally correspond to the likewise underpredicted Einstein radii, and appear in the later test results as well. They do not correlate to any specific lensing configurations, but are instead likely due to poor initialisations in parameter space, and as such do not appear for the CNN and CNN-assisted methods.

For images with LOS structure, results are similar to those above except for a general decrease in accuracy for all results, with errors increasing by factors of 2.2, 8.2 and 2.3 on average for Einstein radius, orientation and axis ratio, respectively. 
This is to be expected as the inclusion of extra mass along the LOS serves generally to increase the complexity of the lensing system compared to smooth parametric profiles, making the resulting images more difficult to model.
As detailed in Section \ref{subsec:testing_sims}, to obtain the 'true' parameters for these images we fitted an SIE model convergence to the convergence of the combined SIE profile and LOS structure.
It is worth noting however that the CNN was not trained on images containing HUDF sources or LOS structure, so while the drop in its accuracy is expected it still performs sufficiently well as an automated modelling method.

The CNN now achieves the highest accuracy for Einstein radii, with errors 10 per cent lower than PyAL + CNN (1$\sigma$), but the latter method continues to give the best results for orientation and axis ratio. For these parameters, the differences between the distributions are reduced compared to the results for images without LOS structure. As such, compared to PyAL (blind) the errors are reduced by 9-28 per cent, 11-17 per cent, and 16-21 per cent for the CNN, PyAL + CNN, and PyAL + CNN (1$\sigma$), respectively.
For the latter method, incorporating the CNN's predicted 1$\sigma$ uncertainties in the priors of \textsc{PyAutoLens} now reduces errors by 3-10 per cent over PyAL + CNN, compared with 13-15 per cent for images without LOS structure.
For both images with and without LOS structure, CNN-predicted uncertainties seen in Figs \ref{fig:predvstrue_sie_hudf_nolos} and \ref{fig:predvstrue_sie_hudf_los} continue to accurately reflect the CNN's errors for the vast majority of lenses, but for those containing LOS structure there are significantly more with greatly underpredicted uncertainties, due to not being represented in the CNN's training set.

\subsection{Power law lenses + HUDF sources}
\label{subsec:pwrlaw_hudf}

\begin{figure*}
  \centering
  \begin{minipage}[t]{\columnwidth}
    \includegraphics[scale=0.44]{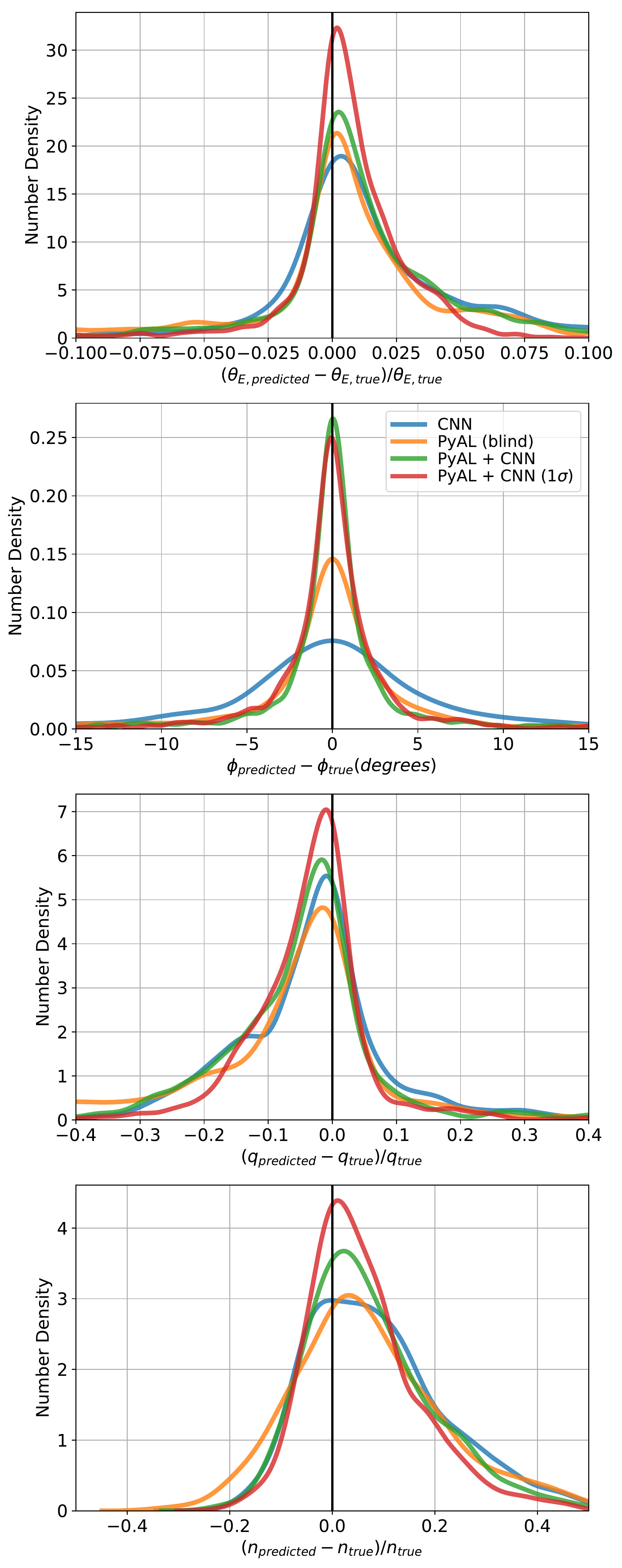}
    \caption{Distribution of the differences between predicted power law mass model parameters and their true values for test data sets of 1000 images containing HUDF sources, without LOS structure.
    The distributions follow the same format as Fig. \ref{fig:kde_sie_parametric}.}
    \label{fig:kde_pwrlaw_hudf_nolos}
  \end{minipage}
  \hfill
  \begin{minipage}[t]{\columnwidth}
    \includegraphics[scale=0.44]{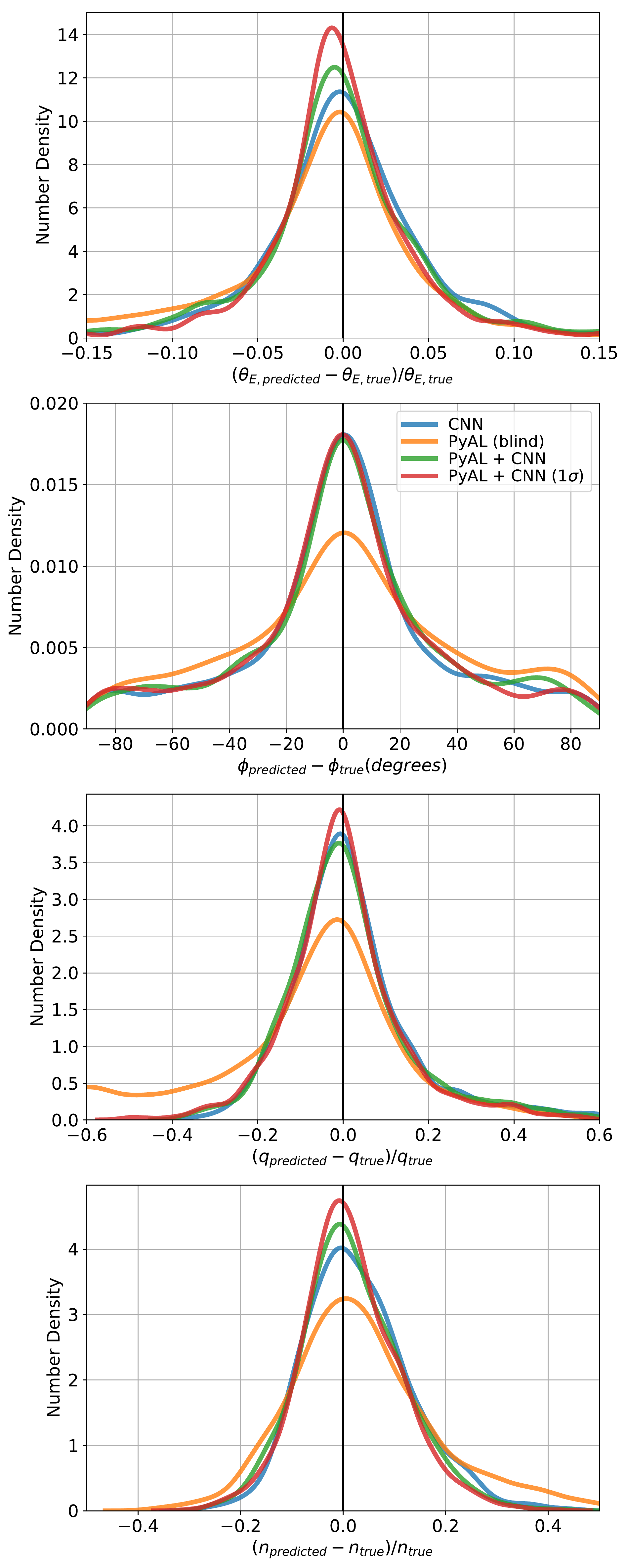}
    \caption{Distribution of the differences between predicted power law mass model parameters and their true values for test data sets of 1000 images containing HUDF sources, with LOS structure.
    The distributions follow the same format as Fig. \ref{fig:kde_sie_parametric}.}
    \label{fig:kde_pwrlaw_hudf_los}
  \end{minipage}
\end{figure*}

\begin{figure*}
  \centering
  \begin{minipage}[t]{\columnwidth}
    \includegraphics[scale=0.44]{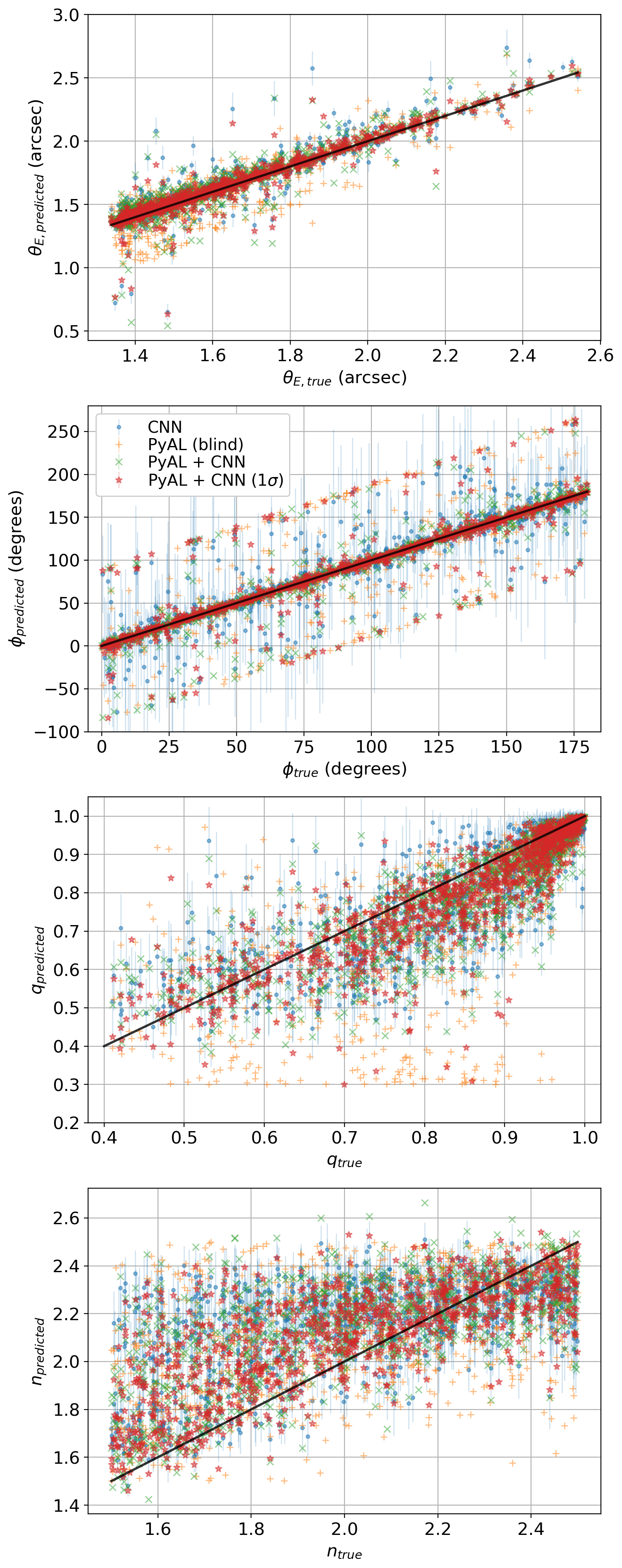}
    \caption{Comparison of predicted power law lens parameters with the true values for test data sets of 1000 images containing power law lenses and HUDF sources, without LOS structure. From top to bottom: Einstein radius, orientation and axis ratio of the lens mass profile.}
    \label{fig:predvstrue_pwrlaw_hudf_nolos}
  \end{minipage}
  \hfill
  \begin{minipage}[t]{\columnwidth}
    \includegraphics[scale=0.44]{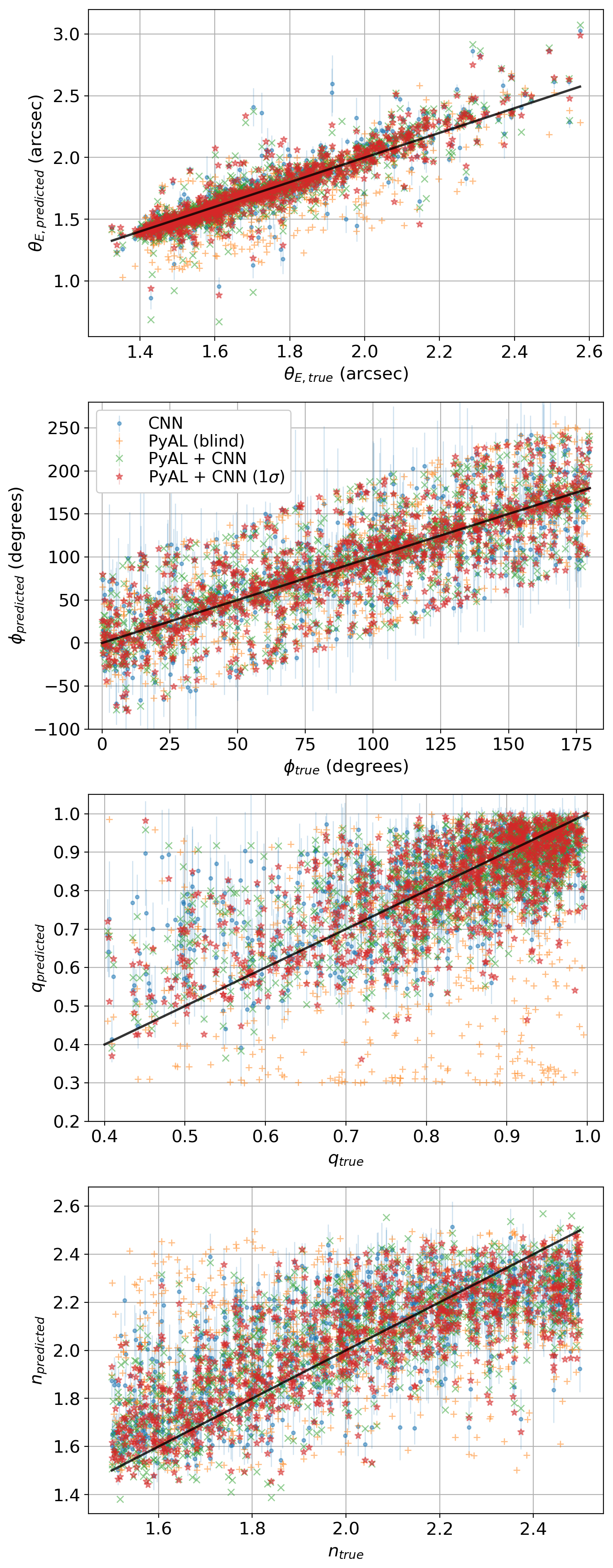}
    \caption{Comparison of predicted power law lens parameters with the true values for test data sets of 1000 images containing power law lenses and HUDF sources, with LOS structure. From top to bottom: Einstein radius, orientation and axis ratio of the lens mass profile.}
    \label{fig:predvstrue_pwrlaw_hudf_los}
  \end{minipage}
\end{figure*}

The same tests as Section \ref{subsec:sie_hudf} were repeated for power law lenses. For this, the same CNN architecture and training process was used but now training and testing on images containing lenses with power law mass profiles instead of SIE profiles. Otherwise the training and testing was identical to that of Section \ref{subsec:sie_hudf}. As such, the power law slope parameter $n$ was also predicted by the network along with its associated 1$\sigma$ uncertainty. 68 per cent confidence intervals are given in Table \ref{tab:pwrlaw_hudf}. Figs \ref{fig:kde_pwrlaw_hudf_nolos} and \ref{fig:predvstrue_pwrlaw_hudf_nolos} present results for images without LOS structure, while figures \ref{fig:kde_pwrlaw_hudf_los} and \ref{fig:predvstrue_pwrlaw_hudf_los} present those containing LOS structure, following the same format as before.

The difficulty of predicting slope values is highlighted in Figs \ref{fig:predvstrue_pwrlaw_hudf_nolos} and \ref{fig:predvstrue_pwrlaw_hudf_los}, with the highest accuracy achieved using PyAL + CNN (1$\sigma$) despite significant scatter remaining in both plots of slope values. Because of this complication, the results for other parameters are generally worse than for images containing SIE lenses, with errors increasing by factors of 1.2-2.9 and 1.1-4.2 for test sets with and without LOS structure, respectively. 
Einstein radii and axis ratios are predominantly affected, especially when modelling with PyAL (blind) whose results contain many more greatly underpredicted axis ratios regardless of the presence of LOS structure.
This is a reflection of the overall degeneracy observed between slope and axis ratio (which in turn affects Einstein radius predictions); for increasing true slope values, predicted slopes in general go from overpredicted to underpredicted, while the opposite occurs for axis ratio predictions. These trends are reversed for increasing true axis ratios.
While the scatter in power law results is larger than that of SIE lenses, the uncertainties predicted by the CNN remain in general an accurate reflection of its errors for both lenses with and without LOS structure. These uncertainties appear to be slightly underpredicted for axis ratios, primarily for the large scatter seen at lower axis ratio values, while uncertainties for slope values appear generally underpredicted.

For the test set without LOS structure, comparisons between modelling methods remain similar to before, with errors of PyAL + CNN (1$\sigma$) 5-39 per cent lower than those of PyAL + CNN, and 32-62 per cent lower than those of PyAL (blind). Meanwhile, the CNN by itself achieves errors comparable to PyAL (blind), from 19 per cent lower to 13 per cent larger. 
With regard to the orientation results seen in Fig. \ref{fig:kde_pwrlaw_hudf_nolos}, the CNN predictions are generally more scattered compared to PyAL (blind). However, while the results of the latter are for the most part centrally concentrated close to the true values, some are clearly offset by around $\pm90^{\circ}$. For PyAL + CNN and PyAL + CNN (1$\sigma$) some of the outlying predictions remain, but these combinations of methods shift the majority of them back towards the true values, decreasing overall errors.

With LOS structure, the CNN and PyAL + CNN (1$\sigma$) methods respectively reduce errors by 23-35 per cent and 27-36 per cent compared to PyAL (blind). Similar to the test set containing SIE lenses and LOS structure, the difference between the CNN and PyAL + CNN (1$\sigma$) methods is smaller than for images without LOS structure, with the latter method obtaining only 0.3-14 per cent lower errors over the former compared to the 28-66 per cent improvement obtained for images without LOS structure. Likewise, the incorporation of 1$\sigma$ uncertainties now only reduces errors by 5-12 per cent over modelling with PyAL + CNN.
The differences in results for images with LOS structure compared to those without are less prominent than those for SIE lenses: now, errors increase by factors of 1.6, 9.2, and 1.3 for Einstein radius, orientation and axis ratio, respectively, while errors in the slope actually decrease by 18 per cent. The modelling methods appear to generally overpredict slope values (i.e. more point-like lenses) in Fig. \ref{fig:predvstrue_pwrlaw_hudf_nolos}, so the addition of LOS structure away from the central lens may allow the modelling to predict sufficiently lower slope values so as to slightly decrease the overall error in these results. The overpredicting of slope values is unlikely to have arisen from differences in their training and testing distributions as both were uniform and sampled between the same bounds.

\begin{table}
	\centering
	\caption{\textbf{Power law lenses + HUDF sources.} The 68 per cent confidence intervals on predicted parameters for each modelling method, computed from the distributions of differences between true and predicted parameter values across 1000 test images.
	The final column contains the confidence intervals for the power law slope parameter.}
	\label{tab:pwrlaw_hudf}
	\begin{tabular}{lllll}
	    \hline
        Method & $\theta_E$ (arcsec) & $\phi$ ($^{\circ}$) & $q$ & $n$\\
		\hline
		{\bf Without LOS structure} & & & & \\
        CNN & 0.050 & 7.37 & 0.083 & 0.27 \\
		PyAL (blind) & 0.065 & 6.53 & 0.106 & 0.29 \\
		PyAL + CNN & 0.049 & 2.63 & 0.080 & 0.25 \\
		PyAL + CNN (1$\sigma$) & 0.029 & 2.50 & 0.062 & 0.20 \\
		{\bf With LOS structure} & & & & \\
        CNN & 0.070 & 32.2 & 0.093 & 0.19 \\
		PyAL (blind) & 0.091 & 44.1 & 0.143 & 0.26 \\
		PyAL + CNN & 0.068 & 34.1 & 0.096 & 0.19 \\
		PyAL + CNN (1$\sigma$) & 0.060 & 32.1 & 0.091 & 0.18 \\
		\hline
	\end{tabular}
\end{table}

\subsection{EAGLE lenses + HUDF sources}
\label{subsec:eagle_hudf}

\begin{figure*}
  \centering
  \begin{minipage}[t]{\columnwidth}
    \includegraphics[scale=0.44]{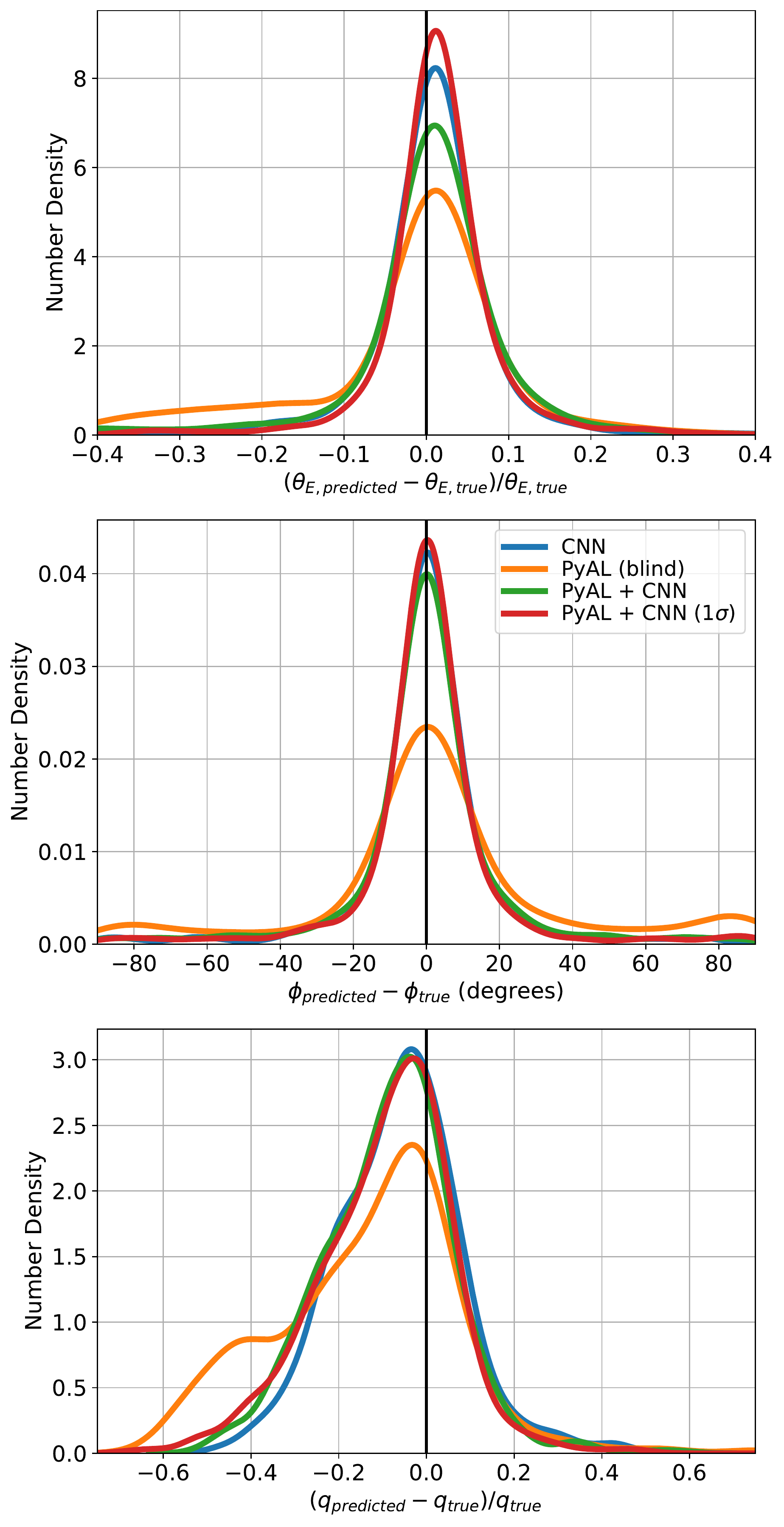}
    \caption{Distribution of the differences between predicted SIE mass model parameters and their true values for test data sets of 1000 images containing EAGLE galaxy lenses with HUDF sources, without LOS structure.
    The distributions follow the same format as Fig. \ref{fig:kde_sie_parametric}.}
    \label{fig:kde_eagle_nolos}
  \end{minipage}
  \hfill
  \begin{minipage}[t]{\columnwidth}
    \includegraphics[scale=0.44]{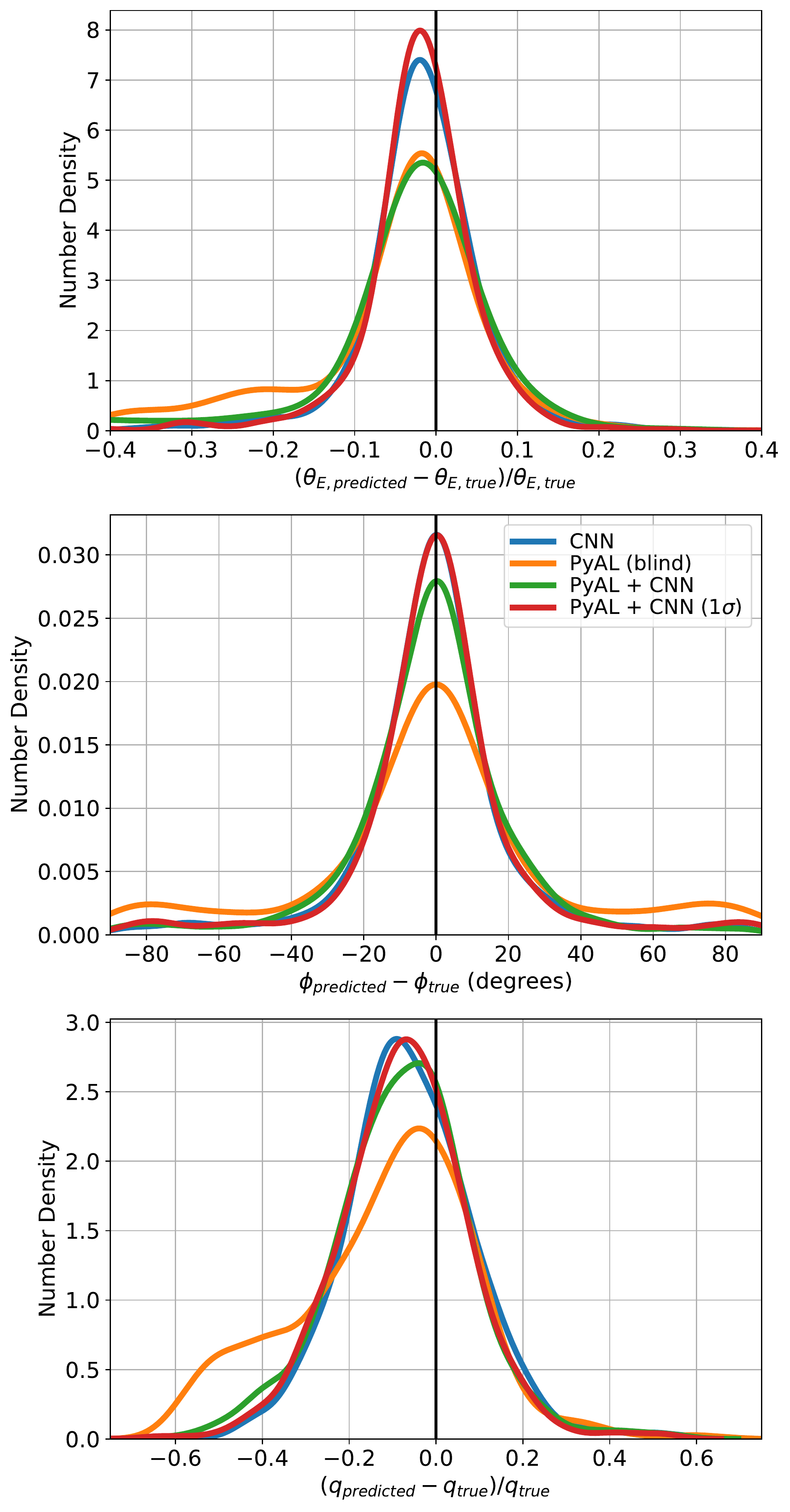}
    \caption{Distribution of the differences between predicted SIE mass model parameters and their true values for test data sets of 1000 images containing EAGLE galaxy lenses with HUDF sources, with LOS structure.
    The distributions follow the same format as Fig. \ref{fig:kde_sie_parametric}.}
    \label{fig:kde_eagle_los}
  \end{minipage}
\end{figure*}

\begin{figure*}
  \centering
  \begin{minipage}[t]{\columnwidth}
    \includegraphics[scale=0.44]{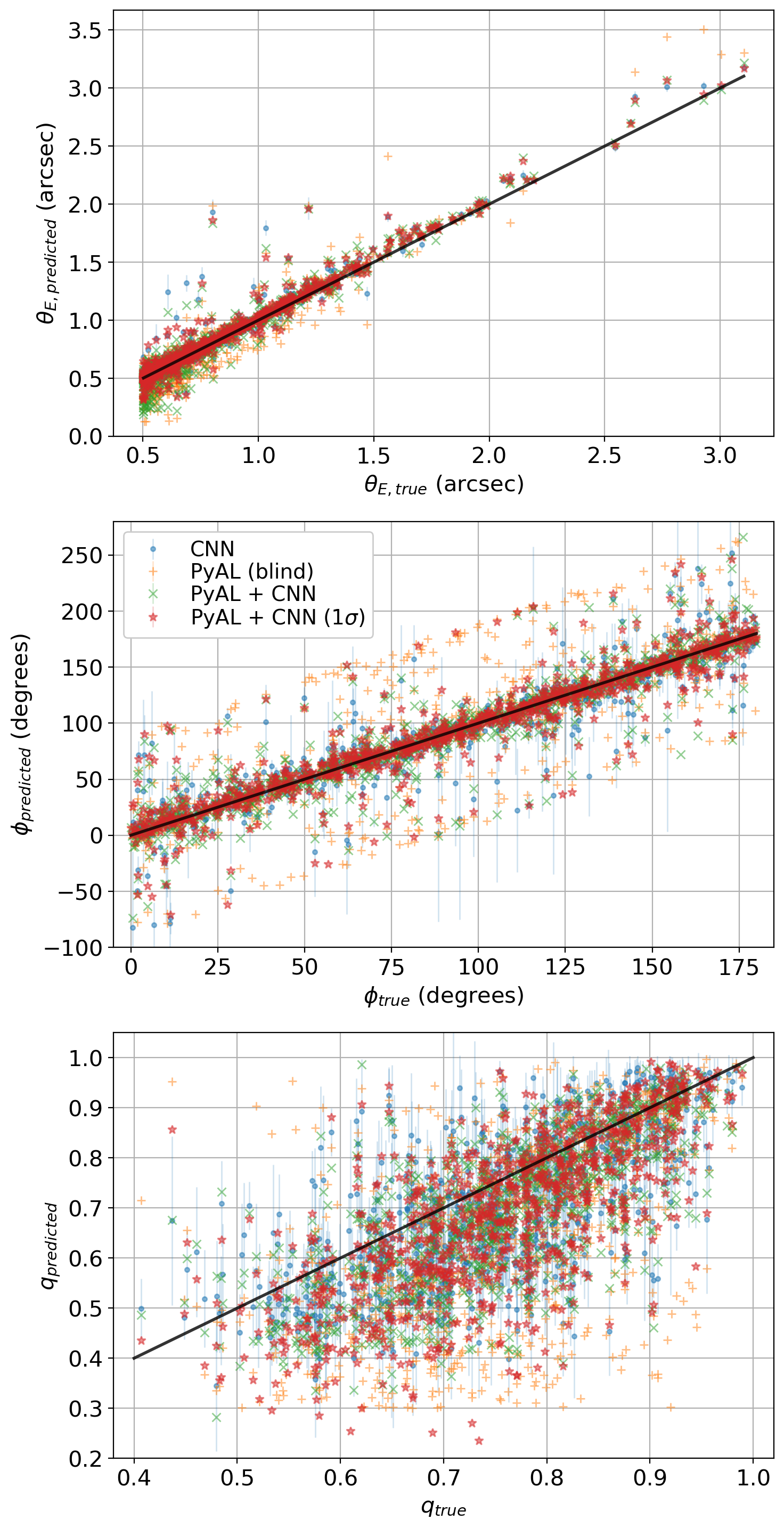}
    \caption{Comparison of predicted SIE lens parameters with the true values for test data sets of 1000 images containing EAGLE galaxy lenses and HUDF sources, without LOS structure. From top to bottom: Einstein radius, orientation and axis ratio of the lens mass profile.}
    \label{fig:predvstrue_siefit_eagle_nolos}
  \end{minipage}
  \hfill
  \begin{minipage}[t]{\columnwidth}
    \includegraphics[scale=0.44]{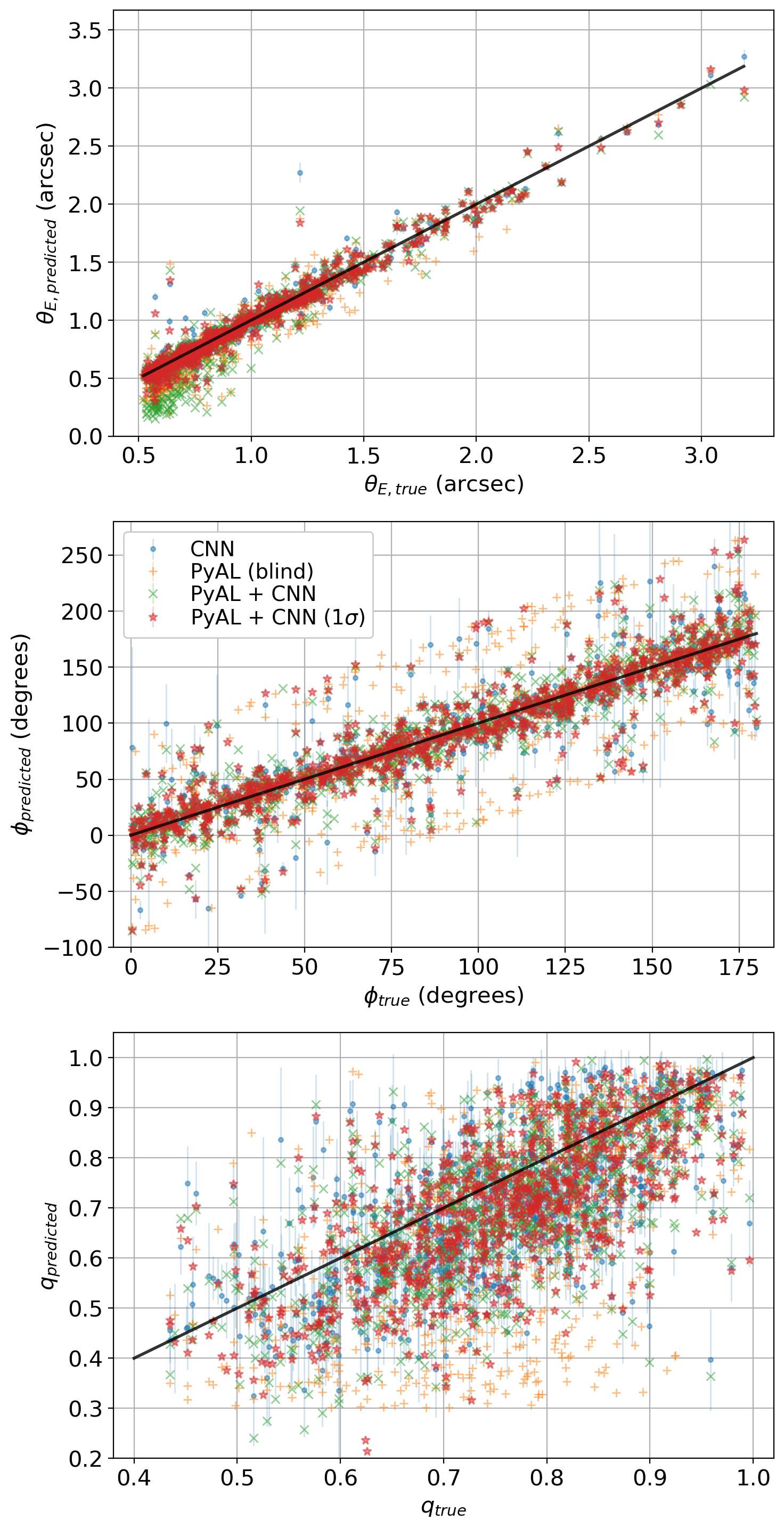}
    \caption{Comparison of predicted SIE lens parameters with the true values for test data sets of 1000 images containing EAGLE galaxy lenses and HUDF sources, with LOS structure. From top to bottom: Einstein radius, orientation and axis ratio of the lens mass profile.}
    \label{fig:predvstrue_siefit_eagle_los}
  \end{minipage}
\end{figure*}

The next stage involved testing on simulated images that contained more complex lenses, i.e. those that do not follow an analytical profile. 
One set of 1000 test images contained lensing galaxies taken from the EAGLE hydrodynamical simulations and the other was identical apart from the addition of LOS structure as applied previously for the parametric lenses.
The known convergence maps of these lenses were fitted with SIE mass profiles in order to obtain the best-fit 'true' parameter values, before modelling the lensed images with an SIE mass profile using the previous methods.
Figs \ref{fig:kde_eagle_nolos} and \ref{fig:kde_eagle_los} show the resulting accuracies of the different modelling methods, with corresponding scatter plots in Figs \ref{fig:predvstrue_siefit_eagle_nolos} and \ref{fig:predvstrue_siefit_eagle_los} and errors presented in Table \ref{tab:eagle_hudf}, in the same format as previous results.

It should be noted that the EAGLE galaxy lenses modelled in this section have smaller Einstein radii on average than in previous sections (see Section \ref{subsec:testing_sims}), producing more pixelated arcs. The modelling methods are therefore expected to produce higher errors for all parameters as a result, on top of the difficulty in modelling more complex mass profiles; for example, see \cite{pearson2019use} for CNN error as a function of Einstein radius.
However, examining the results for lenses with Einstein radii above 1.3 arcseconds, in line with previous test sets, such lenses continue to exhibit much larger errors than those seen in Section \ref{subsec:sie_hudf}, more similar in magnitude to the rest of the EAGLE galaxy data set. As such, it appears the dominant source of error in these results arises from the complex foreground galaxies rather than the smaller Einstein radii.

Without LOS structure, the CNN achieves 36-55 per cent lower errors than PyAL (blind) despite the complex lenses in this test set. The PyAL + CNN (1$\sigma$) method gives similar results, with 30-58 per cent lower errors than PyAL (blind). However, this method no longer outperforms the other CNN methods for all parameters: For Einstein radius and orientation, PyAL + CNN (1$\sigma$) achieves 11-22 per cent and 2-7 per cent lower errors compared to PyAL + CNN and the CNN, respectively, but for axis ratio the PyAL + CNN and CNN methods instead achieve 2.7 and 9.6 per cent lower errors than PyAL + CNN (1$\sigma$).
However, this may simply be due to the large scatter observed in Fig. \ref{fig:predvstrue_siefit_eagle_nolos}, especially  for axis ratio which is often underpredicted by the modelling methods. This is likely caused by two factors: a) the complex foreground galaxies produce fewer smooth, well-defined arcs than those trained on by the CNN, and b) the axis ratios of these galaxies can change as a function of radius, 
with the modelling methods only given final lensed images and hence probe only a part of the whole convergence map used by the separate fitting method to obtain the 'true' values.
Regardless of the presence of LOS structure, both Figs \ref{fig:predvstrue_siefit_eagle_nolos} and \ref{fig:predvstrue_siefit_eagle_los} show that for these complex EAGLE galaxy lenses the CNN underpredicts its uncertainties for all three parameters, with uncertainties barely visible for Einstein radii and clearly not representing the full scatter of the results. As such, were these uncertainties to better represent the errors, PyAL + CNN and PyAL + CNN (1$\sigma$) may have achieved higher accuracies.

Finally, for the test set containing LOS structure, PyAL + CNN (1$\sigma$) achieves the lowest errors for all parameters, but there is again little difference between its and the CNN's results, with PyAL + CNN (1$\sigma$) errors only 0.4-7 per cent lower than the CNN. As such their improvements over PyAL (blind) are almost identical, with PyAL + CNN (1$\sigma$) and CNN errors 28-48 per cent and 28-47 lower than PyAL (blind), respectively. There is still a benefit to incorporating the CNN's predicted 1$\sigma$ uncertainties, which reduce errors by 6-28 per cent over those of PyAL + CNN.
Due to the large amount of scatter in these results caused by modelling the EAGLE galaxy lenses, the presence of LOS structure has only a minor impact on the results, with errors that are factors of 1.3, 1.5 and 1.0 larger than for images without LOS structure for Einstein radius, orientation and axis ratio, respectively.
With LOS structure, errors for the EAGLE galaxy lenses are factors of 2.2, 0.5 and 1.8 larger than those of SIE lenses with LOS structure for Einstein radius, orientation and axis ratio, respectively, compared to factors of 3.9, 3.1, and 4.2 when both test sets do not contain this extra structure.

\begin{table}
	\centering
	\caption{\textbf{EAGLE galaxy lenses + HUDF sources.} The 68 per cent confidence intervals on predicted parameters for each modelling method, computed from the distributions of differences between true and predicted parameter values across 1000 test images.}
	\label{tab:eagle_hudf}
	\begin{tabular}{llll}
        \hline
        Method & $\theta_E$ (arcsec) & $\phi$ ($^{\circ}$) & $q$\\
		\hline
		{\bf Without LOS structure} & & & \\
        CNN & 0.035 & 8.70 & 0.108 \\
		PyAL (blind) & 0.064 & 19.56 & 0.170 \\
		PyAL + CNN & 0.044 & 9.12 & 0.116 \\
		PyAL + CNN (1$\sigma$) & 0.034 & 8.12 & 0.119 \\
		{\bf With LOS structure} & & & \\
        CNN & 0.046 & 12.6 & 0.114 \\
		PyAL (blind) & 0.070 & 23.8 & 0.157 \\
		PyAL + CNN & 0.060 & 15.4 & 0.121 \\
		PyAL + CNN (1$\sigma$) & 0.043 & 12.4 & 0.114 \\
		\hline
	\end{tabular}
\end{table}

\subsection{Modelling speed}
\label{subsec:modelling_speed}

\begin{figure}
  \centering
  \begin{minipage}[t]{\columnwidth}
    \includegraphics[width=\columnwidth]{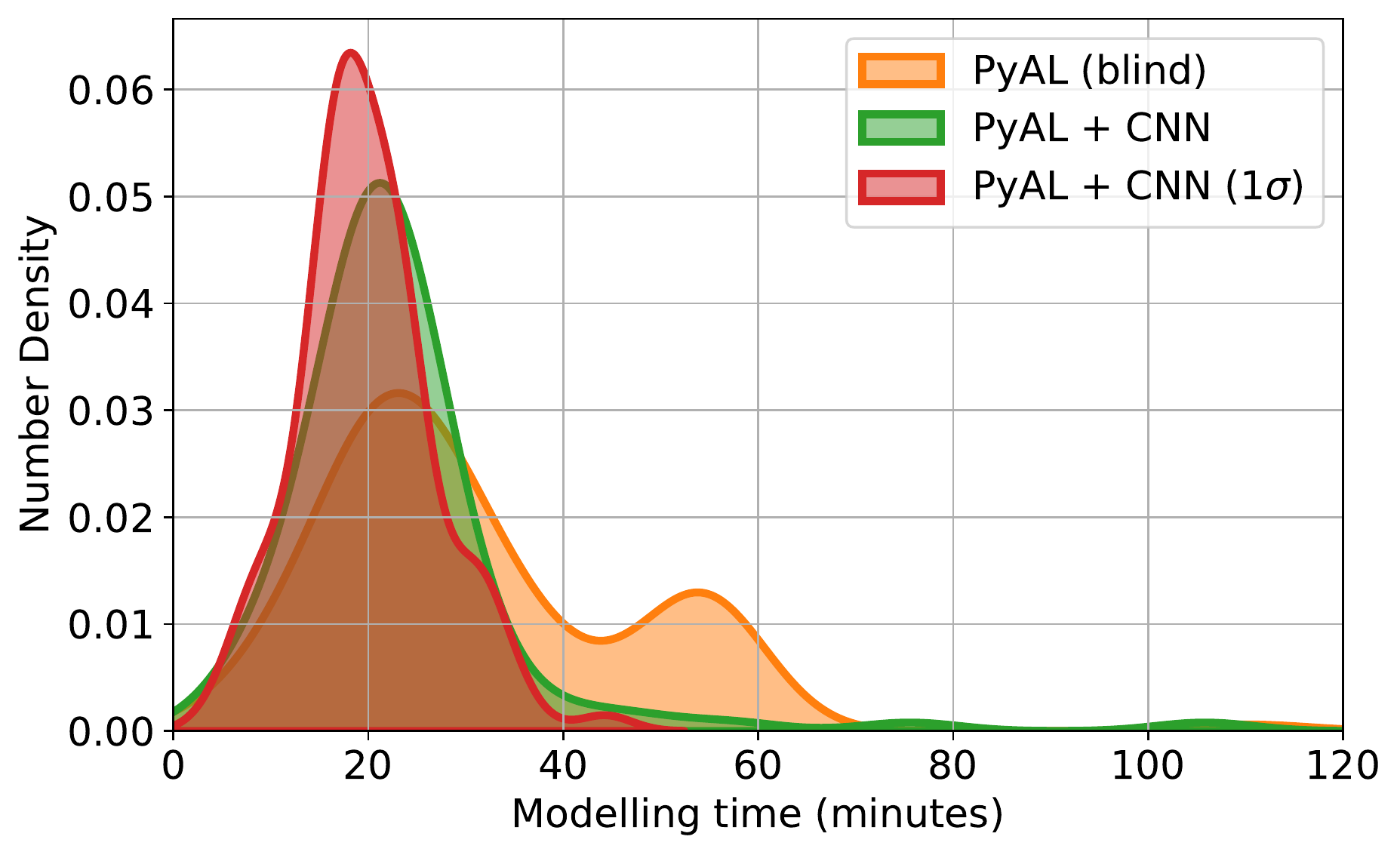}
  \end{minipage}
  \hfill
  \begin{minipage}[t]{\columnwidth}
    \includegraphics[width=\columnwidth]{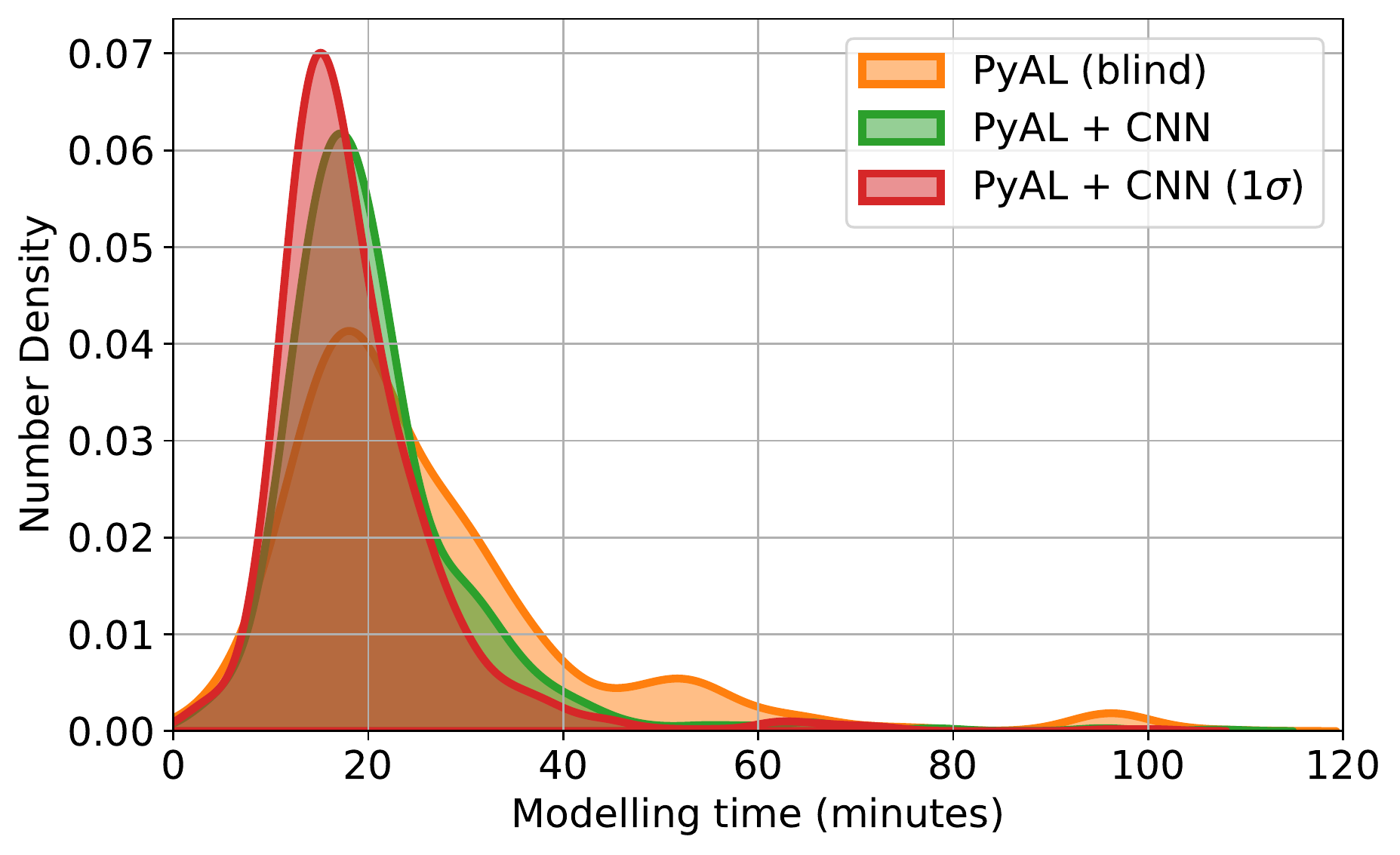}
  \end{minipage}
  \hfill
  \begin{minipage}[t]{\columnwidth}
    \includegraphics[width=\columnwidth]{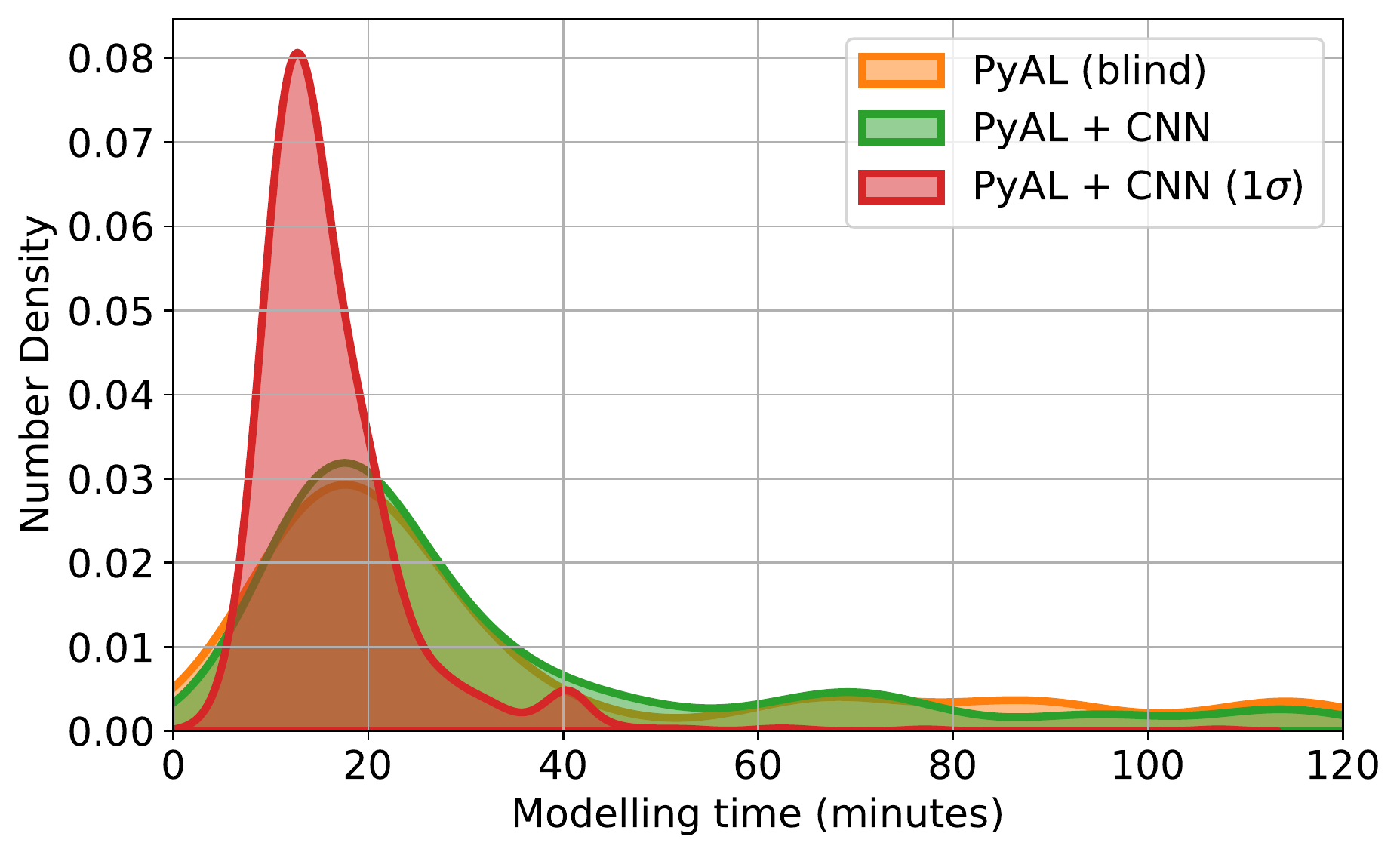}
  \end{minipage}
  \caption{Distributions of the time taken for \textsc{PyAutoLens} to model each of the lenses making up a given test set when modelling blindly, with the CNN predictions as priors, and with the CNN predictions and uncertainties as priors. As the trained CNN can model lenses by itself almost instantly, its modelling times are not included here. From top to bottom: test sets consist of SIE lenses with HUDF sources, SIE lenses with HUDF sources and LOS structure, and EAGLE galaxy lenses with HUDF sources and LOS structure.}
  \label{fig:timings_kde_all}
\end{figure}

We now consider the speed by which the different modelling methods can obtain the lens parameters seen in the previous results. Distributions of the time taken to model each lens using each modelling method are shown in Fig. \ref{fig:timings_kde_all}, for three of the test data sets: SIE lenses with HUDF sources, SIE lenses with HUDF sources and LOS structure, and EAGLE galaxy lenses with HUDF sources and LOS structure. Distributions for the CNN by itself are not shown as for a single lens the time taken is almost instant.

It is clear from the figure that the incorporation of the CNN helps quicken \textsc{PyAutoLens}' modelling for all three test data sets, although nowhere near as fast as the CNN by itself. 
Compared to PyAL (blind), PyAL + CNN increases modelling speeds by mean factors of 1.34, 1.08, and 1.14 for the three test data sets (1.19 on average), while PyAL + CNN (1$\sigma$) increases modelling speeds by mean factors of 1.56, 1.37 and 2.25 (1.73 on average).
A significant improvement is obtained by incorporating the CNN's predicted uncertainties, especially for the EAGLE galaxy lenses; while all methods show some cases of lenses taking a long time to model, PyAL + CNN (1$\sigma$) has far fewer of these outliers, which themselves are generally much shorter than those of the other methods. This improvement arises from applying better priors on parameter space which allow \textsc{PyAutoLens} to converge faster and prevents it from falling into local minima that would give inaccurate parameter values. 
The use of MultiNest by \textsc{PyAutoLens} means that the closer it gets to the correct solution, the greater the number of samples that reach its acceptance threshold. This in turn quickens MultiNest's Markov chain Monte Carlo (MCMC) sampling, further increasing the speed at which \textsc{PyAutoLens} can reach the solution. Hence, starting in more optimal regions of parameter space can greatly quicken modelling.
It is worth noting that the CNN-predicted uncertainties for the EAGLE galaxy lenses were generally underpredicted; while such priors for \textsc{PyAutoLens} may quicken modelling when centred on local or global minima, they are more likely to result in incorrect predictions. Additionally, those not centred on minima would greatly slow the modelling process, and so should more suitable priors be obtained for the EAGLE galaxy lenses, both the accuracy and modelling speed of PyAL + CNN (1$\sigma$) may be increased.

For the automated approach used in this work, only a single run of \textsc{PyAutoLens} is performed for each image, whereas conventionally not only would each image be inspected by eye beforehand, but some modelling may converge on incorrect values and hence require re-initialising with modified priors, further increasing modelling time and the need for human inspection.
The CNN alone provides a much more rapid automated modelling than the combination method, suitable to handle large data sets, while the slower yet still automated PyAL + CNN (1$\sigma$) approach would be well-suited to model more complex lenses for which large uncertainties are predicted by the CNN.

\section{Discussion}
\label{sec:discussion}


We have compared the modelling accuracy of CNNs with that of \textsc{PyAutoLens}' semilinear inversion technique when applied in an automated manner to thousands of simulated images of strong galaxy-galaxy lenses. These images were divided into sets of increasing complexity, from parametric lenses and sources to EAGLE galaxy lenses with HUDF sources and LOS structure, and included modelling both SIE and power law mass profiles. In addition, we examined a promising way to combine both the CNN and \textsc{PyAutoLens} modelling methods, incorporating CNN-predicted parameter values and uncertainties into the priors of \textsc{PyAutoLens}, in order to create an automated method that is faster and more accurate than \textsc{PyAutoLens} would be by itself.

\subsection{Modelling accuracy}


When presented with images containing SIE lenses and parametric sources, all methods reach very high accuracies, with errors below 0.010 arcsecond, 5$^{\circ}$ and 0.03 for Einstein radius, orientation and axis ratio, respectively. This remains the case when the sources are replaced by HUDF galaxies, with errors still below 0.014 arcseconds, 7$^{\circ}$ and 0.04, respectively. The CNN and PyAL (blind) achieve comparable accuracies on average for both of these data sets, however the combination of the two lead to significantly lower errors, with PyAL + CNN (1$\sigma$) errors 37 per cent lower than PyAL (blind) on average across the two data sets. This incorporation of uncertainties improves upon PyAL + CNN by an average of 15 per cent.


Compared to SIE mass profiles, the modelling of power law profiles results in an expected decrease in accuracy, with errors from 0.03 to 0.07 arcseconds, 2.5 to 7.4$^{\circ}$, 0.06 to 0.11, and 0.2 to 0.3 for Einstein radius, orientation, axis ratio and slope, respectively (factors of 1.1 to 4.2 times larger than SIE profile modelling). While the scatter in slope predictions remain large for all modelling methods, there are notable improvements made through the combination of methods, with PyAL + CNN (1$\sigma$) decreasing errors by 5-39, 28-66, and 32-62 per cent compared to PyAL + CNN, the CNN, and PyAL (blind), respectively.

The CNN was not trained on images containing LOS structure, nor to predict external convergence or shear, and hence the introduction of significant LOS structure somewhat complicates the results. Errors increase by factors of 2.2-8.2 for SIE lenses and by 1.3-9.2 for power law lenses, the largest of which are for orientation values, although for the latter images the slope error actually decreases slightly. The differences between PyAL + CNN (1$\sigma$) and the CNN are greatly reduced, however both significantly continue to improve upon PyAL (blind), reaching 27-36 per cent and 23-35 per cent lower errors, respectively.


The similarity between these methods continues for the data set containing EAGLE galaxy lenses. Without LOS structure, PyAL + CNN (1$\sigma$) and the CNN respectively achieve 30-58 per cent and 36-55 per cent lower errors compared to PyAL (blind). With LOS structure, they give almost identical improvements: 28-48 per cent and 28-47 per cent, respectively. As shown, methods involving the CNN continue to give significantly better results than PyAL (blind), despite the CNN not training on such lenses or LOS structure. However, compared to the SIE lenses, errors for all methods increase on average by factors of 3.1 to 4.2, with Einstein radii, orientations and axis ratios respectively predicted with errors of 0.03-0.06 arcseconds, 8-20$^{\circ}$ and 0.10 to 0.17. Given the fairly large scatter as a result of the EAGLE galaxies, the addition of LOS structure has a reduced effect, increasing errors on average by factors of 1.0 to 1.5.

While the CNN was not trained on images containing LOS structure or EAGLE galaxies, such images actually result in the highest improvement over PyAL (blind). This could be because the iterative method that \textsc{PyAutoLens} employs to navigate parameter space is more sensitive to the inclusion of these complex structures than the CNN's instantaneous approach when applied in an automated fashion. The introduction of LOS structure and EAGLE galaxy lenses leads to a decrease in the difference between PyAL + CNN (1$\sigma$) and the CNN. However, due to a lack of sufficient training in these areas, the CNN-predicted uncertainties also appear to be increasingly underpredicted for axis ratios and slopes. Hence, were these to be corrected through further training or otherwise, for example through the incorporation of the unbiased hierarchical inference of \cite{wagner2021hierarchical}, PyAL + CNN (1$\sigma$) could produce a more pronounced increase in accuracy over the CNN for these complex lenses.

\subsection{Modelling speed}


In addition to an increase in accuracy, the combination of methods results in faster modelling times than \textsc{PyAutoLens} by itself. Compared to PyAL (blind), modelling speeds are increased on average by factors of 1.19 and 1.73 
using PyAL + CNN and PyAL + CNN (1$\sigma$), respectively. The priors used in the latter help to reduce the number of outlier lenses that would otherwise take much longer to model, providing a more consistent modelling time. 
It should be noted that for automation, \textsc{PyAutoLens} is only performing a single run for each image; conventionally, human inspection would be used to set these priors, which if incorrect would lead to \textsc{PyAutoLens} falling into local minima and require re-initialising the modelling, further slowing the process. Incorporating CNN predictions therefore takes the place of human inspection, automating the modelling process to deal with upcoming survey data. 

\textsc{PyAutoLens} has a range of configuration options, but while these can lead to obtaining precise fits it can also be time-consuming to choose the best options.
In addition, \textsc{PyAutoLens} requires a high-end PC to run at the speeds shown here. Meanwhile, the CNN need only run on a GPU machine for training, taking less than two hours for 100,000 training images, and while testing the CNN using a GPU machine will speed up modelling, it is not required.
While the neural network alone can be rapidly trained and applied to images, its combination with \textsc{PyAutoLens} provides a slower yet fully automated method of obtaining more accurate results, suitable for applications to lenses that the CNN alone may struggle to model. As such, the CNN would be suitable to handle large samples of upcoming lenses, but should large uncertainties be predicted by the network, the modelling may be passed to PyAL + CNN (1$\sigma$) to refine the predictions.

\subsection{Modelling difficulties and limitations}


The simulated lenses used in Sections \ref{subsec:sie_parametric} to \ref{subsec:pwrlaw_hudf} have a higher average Einstein radius than that expected from lenses detected by \textit{Euclid}. The impact on CNN performance for smaller Einstein radii is detailed in \cite{pearson2019use}, with less available information increasing errors for all parameters. From examining the results in the test sets as functions of Einstein radii we found no significant deviation in the relative accuracies between the modelling methods. Additionally, in Section \ref{subsec:eagle_hudf} the increase in error for the EAGLE galaxy lenses was found to be dominated by these complex foreground galaxies rather than the smaller average Einstein radii. Hence, parametric lens mass profiles with smaller image separations are expected to produce slightly higher errors than for the data sets in Sections \ref{subsec:sie_parametric} to \ref{subsec:pwrlaw_hudf}, but the relative errors between modelling methods would remain consistent.

Compared to SIE profiles, all modelling methods struggle with power law profiles due to the inherent difficulty in accurately measuring slope values, leading to slight biasing towards overpredicting Einstein radii and underpredicting axis ratios.
Results are mostly unbiased for SIE mass profiles, however for a significant number of lenses PyAL (blind) underpredicts Einstein radii and axis ratios, which generally correlate with one another, leading to a large scatter of outliers. These outliers are not present for the other modelling methods, suggesting that the CNN is more reliable in its predictions, which when incorporated into \textsc{PyAutoLens} help it initialise closer to the correct solution in parameter space.

For both mass profiles, the incorporation of significant LOS structure makes modelling challenging, which for this work is equivalent to that of a lens often residing within a galaxy group with many tens of galaxies along the LOS. Replacing the parametric profiles of the lenses with hydrodynamical EAGLE galaxies also serves to increase difficulty, with these complex lenses producing fewer smooth, well-defined arcs than those trained on by the CNN, which can increase the degeneracy between axis ratio and orientation when modelling \citep{mukherjee2018seagle}. These lenses also in general have smaller Einstein radii than the parametric lenses, corresponding to more pixelated arcs and reduced information available for modelling.
Additionally, the large errors seen in axis ratios for the EAGLE galaxy lenses can be attributed to how these values often change as a function of radius. This, along with many lenses containing significant substructure in their convergence maps, especially when combined with LOS structure, prove difficult to reliably fit in order to obtain the parameter values used as "true" values for this work.

One aspect of consideration in this work is the source-lens alignment, i.e. the distribution of doubles and non-doubles (quads, rings) within the data sets. While doubles are expected to dominate the catalogues of future surveys, such images are in general more difficult to model. In this work, a lower limit on magnification was used to ensure lensed images were being generated; this focused on sources in higher regions of magnification, resulting in only around 30 to 40 per cent of test images containing doubles. As such, we examined how the performance of the modelling methods varied by analysing 100 doubles and 100 non-doubles in the test set containing SIE lenses and HUDF sources. On average across the modelling methods, errors in Einstein radii and axis ratios for doubles were factors of 2.1 and 1.9 times larger than non-doubles (1.5 and 1.3 times larger than the combined 200-image data set), respectively. Meanwhile, orientation errors for doubles were on average a factor of 0.8 times that of non-doubles (and were almost unchanged compare to the combined data set), likely because the set of non-doubles contained near-complete Einstein rings. However, while the modelling errors differed between doubles and non-doubles, the relative accuracies between the modelling methods remained broadly similar.


While the image data sets used throughout this work have contained only lensing systems, future real data sets may inadvertently contain non-lenses if the image classification process is not perfected. The expectation is that attempting to model non-lenses with \textsc{PyAutoLens} will produce poor evidence values which would indicate a problem with the modelling.
However, the uncertainties predicted by the CNN may be of little use in this regard; while it is possible that they may increase for non-lenses due to a lack of a preferred lensing model, the CNN was not trained on such images and so their impact remains unknown without further testing. This, along with re-training the CNN on both lenses and non-lenses to ensure appropriate predicted uncertainties, is beyond the scope of this work and instead left for future investigation.

Likewise, the performance of the modelling methods may be impacted by how well light is masked or subtracted from the lens and companions in the field of view, which would require future work; for example, methods of denoising and deblending lens and source light have been developed by \cite{madireddy2019modular}. The test images here assumed perfect light subtraction for all but the background source galaxy, although the impact of CNN performance when tested on images containing lens light was investigated in \cite{pearson2019use}. PyAutoLens has been designed to accommodate masked images, while much of the impact of masking on CNN performance would be remedied through appropriate training. If residuals were present from incorrect masking or subtraction, PyAutoLens may struggle to fit the source light profile, as would the CNN if not trained on such imperfections. However, given sufficient CNN training in these areas, it is likely that the conclusions presented here would remain approximately unchanged.

\section{Summary}
\label{sec:summary}


Strong gravitational lensing can be used to study galaxy evolution, probe high-redshift source populations and constrain cosmological models. However, the complicated process of modelling these lenses has previously necessitated the use of techniques such as the parametric parameter-fitting of \textsc{PyAutoLens}, which are often relatively slow and require manual inspection.
With the advent of large-scale surveys soon to observe tens of thousands of strong lenses, automated techniques will be required to model these lenses quickly and efficiently. As a result, this has driven the use of machine learning, especially CNNs, to recover lens mass model parameters extremely quickly once trained.


In this work, we trained an approximate Bayesian CNN to predict strong gravitational lens mass model parameters and compared its accuracy to that of the semilinear inversion technique of \textsc{PyAutoLens}, when applied automatically to a range of increasingly complex lensing systems. These included parametric SIE and power law lens mass profiles, EAGLE galaxy lenses, and line-of-sight structure. In addition, we also combined these modelling methods together, with CNN predicted values and uncertainties acting as priors for \textsc{PyAutoLens}.


Across the data sets and lens parameters, the CNN errors are 19 $\pm$ 22 per cent lower than PyAL (blind) on average. Hence, for the majority of cases, the CNN results are on par with or exceed those of \textsc{PyAutoLens}, whose modelling is not typically automated, instead requiring manual inspection of images to obtain suitable priors.
The combination of these two methods when using just CNN predicted values as priors, PyAL + CNN, improves upon PyAL (blind), achieving 27 $\pm$ 11 per cent lower errors across the parameters. However, it is frequently matched or exceeded by the CNN, with average errors only 4 $\pm$ 23 per cent lower than the network. While the CNN is by far the quickest modelling approach, this combination enhances the modelling speed of \textsc{PyAutoLens}, increasing it by a factor of 1.19 over PyAL (blind).

A substantial improvement is obtained when the CNN-predicted uncertainties are also incorporated into \textsc{PyAutoLens}' priors, PyAL + CNN (1$\sigma$). These uncertainties act to better constrain the prior distributions which help \textsc{PyAutoLens} to avoid local minima and converge faster on the correct solution. With these included, this combination of the two methods produces significantly higher accuracies than either one separately: PyAL + CNN (1$\sigma$) reduces errors on average by 37 $\pm$ 11 per cent across the parameters compared to PyAL (blind), 17 $\pm$ 21 per cent compared to the CNN, and 13 $\pm$ 9 per cent compared to PyAL + CNN. Additionally, PyAL + CNN (1$\sigma$) outperforms PyAL (blind) in all tests, as well as the CNN in the majority of cases. It also gives the highest \textsc{PyAutoLens} modelling speeds, increasing upon PyAL (blind) by a factor of 1.73 on average across the test sets. It should be noted that for the more complex lensing systems these uncertainties are often underpredicted, so should these be corrected an even greater improvement may be achieved.


The CNN's modelling speed makes it well-suited for large catalogues of lenses, and given the many thousands of upcoming strong lens observations such a network can very quickly predict accurate values and uncertainties for the majority of cases. While its accuracy in this work can rival that of the combination method for some lenses, it is limited by the complications and artefacts encountered in real data. Until the CNN can be trained on sufficiently realistic \textit{Euclid} survey data, the PyAL + CNN (1$\sigma$) combination provides a promising alternative automated modelling method. Those challenging for the CNN to model produce larger CNN uncertainties, which can then be passed to \textsc{PyAutoLens} to verify and refine predictions. This combination of methods acts as a fully automated pipeline that can achieve accurate results far quicker than conventional modelling. Rather than keeping them separate, this work highlights the importance of considering the combination of machine learning with conventional approaches in order to gain the benefits of both.



\section*{Acknowledgements}
\label{sec:acknowledgements}


We thank the anonymous referee for their helpful comments and suggestions. We also greatly appreciate the support and contributions made by Tony and Alex Pearson.
JM and NL acknowledge the support by the UK Science and Technology Facilities Council (STFC).
SD is supported by a UK STFC Rutherford Fellowship.



\section*{Data Availability}
\label{sec:data_availability}


The data supporting this work may be shared upon reasonable request to the corresponding author.




\bibliographystyle{mnras}
\bibliography{references} 








\bsp	
\label{lastpage}
\end{document}